\documentclass[%
 reprint,
superscriptaddress,
 amsmath,amssymb,
 aps,
prl,
]{revtex4-2}
\usepackage{physics}
\usepackage{graphicx}
\usepackage{dcolumn}
\usepackage{bm}
\usepackage[colorlinks=true,linkcolor=blue,urlcolor=blue,citecolor=blue]{hyperref}

\usepackage{soul}
\usepackage{lipsum}  
\setcitestyle{super} 

\usepackage{newtxtext, newtxmath}

\begin{document}


\title{Compute-first optical detection for noise-resilient visual perception}

\author{Jungmin Kim}
\email{jmkim93@gmail.com}
\affiliation{Department of Electrical and Computer Engineering, University of Wisconsin-Madison, Madison, WI 53706, USA}

\author{Nanfang Yu}
\affiliation{Department of Applied Physics and Applied Mathematics, Columbia University, New York, NY 10027, USA}

\author{Zongfu Yu}%
\email{zyu54@wisc.edu}
\affiliation{Department of Electrical and Computer Engineering, University of Wisconsin-Madison, Madison, WI 53706, USA}

\date{\today}

\begin{abstract}
In the context of visual perception, the optical signal from a scene is transferred into the electronic domain by detectors in the form of image data, which are then processed for the extraction of visual information. In noisy and weak-signal environments such as thermal imaging for night vision applications, however, the performance of neural computing tasks faces a significant bottleneck due to the inherent degradation of data quality upon noisy detection. Here, we propose a concept of optical signal processing before detection to address this issue. We demonstrate that spatially redistributing optical signals through a properly designed linear transformer can enhance the detection noise resilience of visual perception tasks, as benchmarked with the MNIST classification. Our idea is supported by a quantitative analysis detailing the relationship between signal concentration and noise robustness, as well as its practical implementation in an incoherent imaging system. This compute-first detection scheme can pave the way for advancing infrared machine vision technologies widely used for industrial and defense applications.
\end{abstract}

\maketitle


\section{Introduction}
Recent advances in infrared (IR) technologies around atmospheric windows have expedited various scientific and industrial fields, including night vision technologies based on thermal imaging \cite{Rogalski2010, Yang2021, Bao2023, Marnissi2023} and radiative cooling systems addressing the global climate crisis \cite{Raman2014, Yang2021, Fan2022, xZhao2023, Munday2019}, which use infrared light as an information and heat carrier, respectively. These technologies commonly leverage transmission within the mid-IR regime, relying on blackbody radiation \cite{Rogalski2010, Raman2014} emitted from an object at around room temperature without external sources. However, the relatively weak IR power, compared with that of the daytime ambient light, has posed a challenge: the low signal-to-noise ratio (SNR) in thermal imaging in the presence of detection noise. Several studies have focused on the post-processing of noisy images to overcome the low SNR issue by incorporating additional degrees of freedom such as hyperspectral \cite{Bao2023, Amenabar2017, Wang2019, Zhao2023, xWang2023} or polarimetric \cite{Gurton2014, Yuffa2014, Deng2022} information, which involved developing apparatus for the fast acquisition and processing of large datasets. 

Regarding visual perceptions \cite{Szeliski2022} such as object recognition and feature detection from a noisy environment, plenty of additional computing mechanisms in the optical domain based on diffractive \cite{Lin2018, Wu2019, Wu2021, Bernstein2023, yChen2023, Fu2023, TWang2023, Zheng2023} or interferometric \cite{Shen2017, Hamerly2019, MourgiasAlexandris2022, ZChen2023, Zheng2023, Yu2023} devices can be employed to resolve this issue. The basic idea is to focus on how to obtain cleaner data instead of on how to better deal with noisy data using digital post-processing. This is inspired by Fourier-transform infrared (FTIR) spectrometer in comparison with a grating-based monochromator \cite{Colthup1990}. While the monochromator spatially separates each spectral component using a diffraction grating mirror to capture the spectral information of light, the FTIR interferometer does not scatter light but rather encodes the spectral information into a temporal pattern with a higher SNR, allowing for the computational decoding of the spatially condensed signals using Fourier transformation (FT) afterwards.

In the traditional approach to machine visual perception, firstly one needs to obtain the image of a scene by imaging devices. As displayed in Fig. 1a, the wave signal of the image is then transferred to an electronic domain by “detection” with a photodetector (PD) array, and the useful visual information (i.e., the feature) of the scene is extracted from the acquired image data through a series of data processing procedures. However, this detection-computing sequence places its computational load fully behind the detection, resulting in the inherent vulnerability to noises, such as the thermally generated dark current in PDs \cite{Rogalski2010}. In this work, we aim to leverage additional optical computing mechanisms or resources to address this issue, as depicted in Fig. 1b: enhancing SNR with pre-detection optical processing unit (OPU) that is capable of concentrating the optical signal without loss of information. To validate the idea, we demonstrate a theoretical framework where a properly designed optical neural network is integrated with a digital MNIST \cite{Cohen2017} classifier, revealing the enhanced performance in terms of the resilience of classification accuracy against an extreme dark noise. Quantitative evidences are then provided to establish the relation between the robustness and the degree of concentrative modulation along with the concept of detection pruning. Finally, we demonstrate an incoherent imaging system as a practical example, verifying the superior robustness against noise with the data-driven design of a metalens system.

\begin{figure}[t!]
    \centering
    \includegraphics{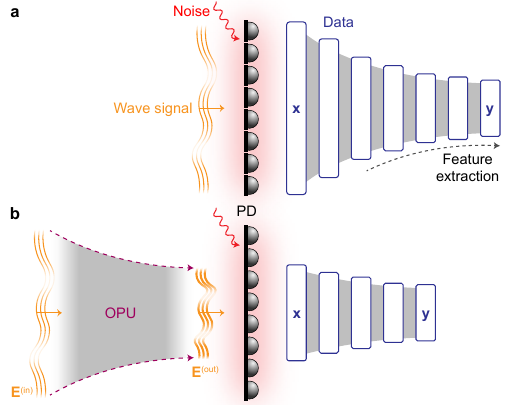}
    \caption{\textbf{Concept of optical compute-first detection system for visual perception}. \textbf{a}, Conventional procedure: the wave signal from a scene is converted to image data by a photodetector (PD) array, with additional detection noise. Subsequently, a digital processor processes the image data, extracting a latent feature of the scene.  \textbf{b}, Proposed scheme: the wave signal undergoes primary modulation ahead of detection through an optical processing unit (OPU). It is detected and then post-processed in the digital domain to produce the final visual information. $\vb E^\mathrm{(in,out)}$, input and output state of waves; $\vb x$, detected value in electronic domain; $\vb y$, target feature.} 
    \label{fig1}
\end{figure}

\section{Results}

\subsection{Model definition}
In the traditional approach to visual perception tasks, one needs first to obtain the image of a scene, which is regarded as a tailored copy of a scene at a detection plane by imaging devices. As displayed in Fig. 1a, the replicated wave signal is then transferred to a digital domain by ``detection'' with a photodetector array, and then the acquired image data results in the perception of the scene through neuromorphic image processing. However, this imaging-and-processing sequence places its computational load fully after detection, leading to inherent noise issues. Our objective is to leverage additional optical computing resources before detection, as depicted in Fig. 1b, to enhance the SNR.

We assume that the OPU depicted in Fig. 1b operates as a linear system based on the superposition principle for the electric field. Therefore, we can streamline such linear optical devices through discretization: $\vb E^\mathrm{(out)} = f(\vb E^\mathrm{(in)}; P) = P \vb E^\mathrm{(in)}$, where the input and output vectors $\vb E^\mathrm{(in, out)} \in \mathbb C^{1\times N}$ have finite spatial dimensions N. Further assuming that the total energy is conserved during optical signal processing: $\ip*{\vb E^\mathrm{(in)}}{\vb E^\mathrm{(in)}} = \ip*{\vb E^\mathrm{(out)}}{\vb E^\mathrm{(out)}}$, the transfer matrix $P$ should be unitary, i.e., $P^\dagger P = I$. We note that this discrete and unitary constraint aligns with the solution of the coupled-mode equation for a waveguide system \cite{Carolan2015}. Importantly, any arbitrary unitary operation can be programmed using Mach-Zehnder interferometers and phase shifters with the same degrees of freedom ($N^2$). Well-known Clements \cite{Clements2016} and Reck \cite{Reck1994} designs serve as effective tools for achieving this programmability. Hence, a discrete unitary system emerges as an effective testbed for the analysis and demonstration of pre-detection optical processing.

As a representative task for machine visual perception, we benchmark the MNIST classification performance \cite{Cohen2017} using two cascaded networks: a deep neural network as a digital processor attached to the linear OPU, as illustrated in Fig. 1b. The digital network, $\vb y = g(\vb x; Q)$, performs the post-processing of the optical intensity signal $\vb x \in \mathbb{R}^{1\times N}$ with a trainable parameter set $Q$ to infer the visual feature: in this case, the digital network generates the probability distribution $\vb y \in \mathbb{R}^{1\times M}$ over $M$ classes, by which a decision can be made to the most-probable class. Specifically, our target task is the classification of MNIST objects with $28\times 28$ resolution, therefore, we have $N = 28^2$ and $M = 10$.

Meanwhile, there is assumed to be a physical detection process (i.e., a transition from an optical to an electronic signal) over PDs \cite{Rogalski2010} between the two domains. The PD array typically measures the photon counts incident to each pixel, which is a function of the output intensity vector $\vb x = h(\vb E^\mathrm{(out)})$ with element-wise operations:
\begin{equation}
    x_\alpha = |E_\alpha^\mathrm{(out)}|^2 + \Delta I_\mathrm{photon} + \Delta I_\mathrm{dark},
\end{equation}
where $\alpha = 0, \cdots, N – 1$ is the pixel index, and
\begin{align}
    \Delta I_\mathrm{photon} &\sim \frac{\mathrm{Pois}(\Delta t |E_\alpha^\mathrm{(out)}|^2 )}{\Delta t} - |E_\alpha^\mathrm{(out)}|^2, \mathrm{~and~} \\
    \Delta I_\mathrm{dark} &\sim N(0, \sigma_\mathrm{dark}^2)
\end{align}
represent two independent noise mechanisms typically involved in the optoelectronic detection: $\Delta I_\mathrm{photon}$ is photon shot noise, a Poisson random process arising from the discrete nature of photons arriving at each detector within a time frame $\Delta t$; $\Delta I_\mathrm{dark}$ accounts for all other input-independent noises such as thermal and dark current noise, approximated by a Gaussian process with effective noise power $\sigma_\mathrm{dark}$.

\begin{figure*}[t!]
    \centering
    \includegraphics{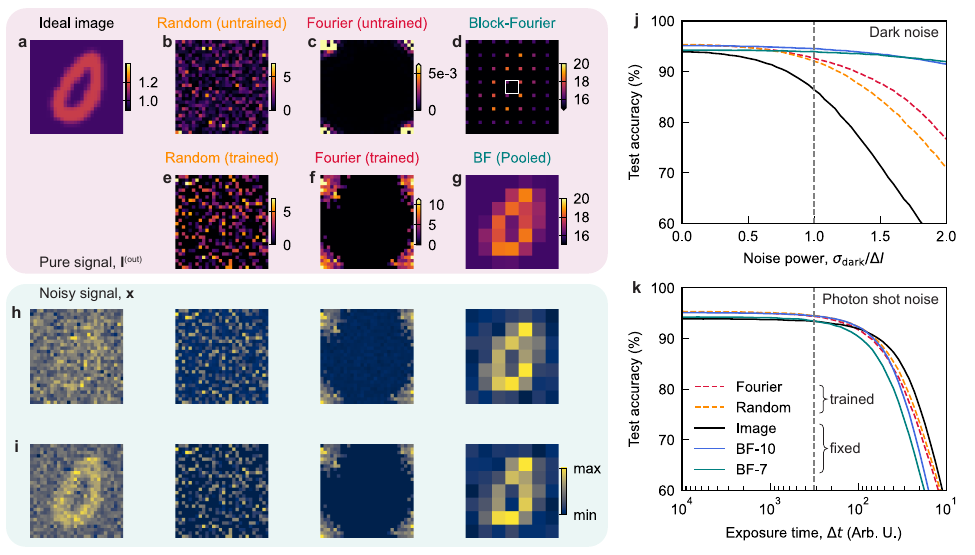}
    \caption{\textbf{Noise robustness achieved by optical signal processing.} \textbf{a}-\textbf{g}, 2D representations ($28^2$ pixels) of optical intensities before detection, $I_\alpha^\mathrm{(out)}$: ideal image of digit 0 (\textbf{a}), random matrix multiplied image (\textbf{b}), 2D Fourier image (\textbf{c}), block-wise 2D Fourier image (\textbf{d}), images with machine-trained unitary matrices (\textbf{e} and \textbf{f}) from the initialization with \textbf{b} and \textbf{c}, respectively, and sampled ($7^2$ pixels) image from \textbf{d} by max-pooling (\textbf{g}). \textbf{h},\textbf{i}, Detected images with two different types of noise $x_\alpha = I_\alpha^\mathrm{(out)} + \Delta I_\mathrm{dark} + \Delta I_\mathrm{photon}$: dark noise (\textbf{h}, $\Delta I_\mathrm{photon}\sim 0$) and photon shot noise (\textbf{i}, $\Delta I_\mathrm{dark}= 0$), applied to \textbf{a}, \textbf{e}, \textbf{f}, and \textbf{g} from left to right. \textbf{j}, \textbf{k}, MNIST classification accuracies according to increasing test noise levels: dark noise power (\textbf{j}) and shot exposure time (\textbf{k}), for various optical processing types (ideal image \textbf{a}, black; machine-trained operations \textbf{e} and \textbf{f}, red and orange; fixed block-wise Fourier operations \textbf{g} with different segmentation numbers 10 and 7, blue and green, respectively). Grey dashed lines indicate the applied noise level in \textbf{h} and \textbf{i}. The test accuracy is calculated over $10^4$ balanced test samples with 20 repetitions. $\Delta I\sim0.17$ is the intensity contrast in ideal images (\textbf{a}).}
    \label{fig2}
\end{figure*}

\subsection{Noise-resilience of compute-first detection scheme}

To demonstrate the robustness against detection noise achieved by the optical pre-processing, we investigate two different types of linear OPUs (i.e., designed $P$). Firstly, $P$ can be trained as part of the total parameter set of tandem optical-digital networks $\vb y = (g \circ h \circ f)(\vb E^\mathrm{(in)}; P, Q)$, as demonstrated by deep learning \cite{Lin2018, Wu2019, Wu2021, Rahman2023} or the adjoint-based optimization \cite{Piggott2015, Hughes2018, Khoram2019} of optical elements, although it can converge into different local optimal solutions with the stochastic gradient-descent method, depending on how it is initialized. Otherwise, $P$ can be assigned a manually defined unitary matrix that is likely to concentrate the optical signal. On the contrary, we can set $P$ as the identity matrix for the reference model, representing ideal imaging devices without proper optical treatment. For all cases, we optimize the digital network $g$ through supervised learning, employing cross-entropy loss
\begin{equation}
    L(\vb y, \vu y) = -\sum_{m=0}^{M} y_m\log \hat{y}_m
\end{equation}
for a one-hot encoded class label $\vu y$. Further details on the model architecture are available in Supplementary Table S1 and Fig. S1, and learning curves are referred to in Fig. S2.

Figure 2a, for instance, shows the coherent input intensity distribution (and the identical output image for the reference model) $I_\alpha^\mathrm{(in)} \equiv |E_\alpha^\mathrm{(in)}|^2$ of Class 0, 2D-reshaped into $28^2$ pixels. This input can be processed by a random-generated unitary matrix $P_\mathrm{R}$ or the 2D discrete Fourier transform (DFT) matrix $P_\mathrm{F}$, resulting in the output intensity distributions $I_\alpha^\mathrm{(out)} = |E_\alpha^\mathrm{(out)}|^2$ (Figs. 2b and 2c, respectively). Alternatively, using a manually designed block-wise Fourier matrix $P_\mathrm{BF,7}$, which divides the domain into several rectangular blocks and operates DFT in each block (see Methods section for detailed definition) and is inspired by the micro-lens array structure \cite{Lin2019}, the output intensity distribution can be focused into several representative pixels (Fig. 2d; white border indicates one of the square blocks). Since the results in Figs. 2b and 2c are not yet optimized, we further train $P_\mathrm{R, F}$ through deep learning to $\tilde{P}_\mathrm{R, F}$, resulting in the output intensity distributions optimal for the following neural inference as depicted in Figs. 2e and 2f, respectively. Simultaneously, we apply max-pooling, i.e., dimensionality reduction by taking the maximum value for each subdomain, to the block-wise Fourier result (Fig. 2g) to transfer only the DC component, i.e., maximum-intensity pixel per block, to the subsequent inference $g$. It is noteworthy that both the machine-optimized (Figs. 2e and 2f) and manually defined (Fig. 2g) OPUs effectively concentrate the input signal distribution (Fig. 2a), allowing for the enhancement in intensity contrast by an order of magnitude up to $10^1$, compared to the input intensity contrast $\Delta I \equiv \max_\alpha(I^\mathrm{(in)}_\alpha)-\min_\alpha(I^\mathrm{(in)}_\alpha) \sim 0.17$, which is chosen based on the black-body radiation contrast within a temperature range from 300 to 310 K for a LWIR wavelength ($10~\mu\mathrm{m}$).

Applying the dark noise $\Delta I_\mathrm{dark}$ and the photon shot noise $\Delta I_\mathrm{photon}$ independently upon optical-to-electronic transition, we can observe the capability of such optical treatments in compensating unavoidable dark noise. For instance, the left to right inset of Fig. 2h displays the detected signals $\vb x$ with dark noise of noise power $\sigma_\mathrm{dark} = \Delta I$ for the ideal image (Fig. 2a) and OPU output (Figs. 2e-2g), respectively. The reference result (left, Fig. 2h) is almost masked by the strong dark noise, making it challenging to identify as the digit ``0". In sharp contrast, the block-wise Fourier result (right, Fig. 2h) can be interpreted as Class 0, despite its coarse mosaic effect, due to the magnified output intensity contrast.

This distinct difference between dark noise-screened signals can be analyzed through a quantitative MNIST benchmark. First, we train the combined optical-digital networks with the fixed degree of noise levels: $\sigma_\mathrm{dark}^\mathrm{(tr)}=\Delta I/\sqrt{2}$ and $\Delta t^\mathrm{(tr)}=2/\Delta I^2$, thus the effective noise power being $\sigma_\mathrm{eff}^\mathrm{(tr)} = [(\Delta t^\mathrm{(tr)})^{-1}+ (\sigma_\mathrm{dark}^\mathrm{(tr)})^2]^{1/2}=\Delta I$ for a unit intensity. Then, we test the trained networks with an increasing dark noise level $\sigma_\mathrm{dark}$ from zero, as shown in Fig. 2j. The green and blue solid lines for the manually designed block-Fourier matrices with 7 segments ($P_\mathrm{BF,7}$) and 10 segments ($P_\mathrm{BF,10}$), respectively, exhibit extreme robustness against dark noise up to $\sigma_\mathrm{dark}\sim 2\Delta I$. In contrast, the test accuracy for the ideal image without optical processing (black line) rapidly decreases with dark noise. The machine-optimized models with different initialization ($\tilde{P}_\mathrm{F}$ and $\tilde{P}_\mathrm{R}$; red and orange dashed lines) also outperform the reference model.

Interestingly, the linear OPUs are not effective for photon shot noise in enhancing SNR, as the absolute noise power of the shot noise is simultaneously amplified when the signals are concentrated, as $\Delta I_\mathrm{photon} \propto [I_\alpha^\mathrm{(out)}]^{1/2}$. That is, the shot noise is more related to the total computing energy per operation as in Ref.\cite{Hamerly2019}. This difference is evidenced by the noisy images in Fig. 2i (ideal image on the left, machine-trained in the middle, and block-Fourier on the right), as well as by Fig. 2k illustrating almost no difference in noise robustness of various models with decreasing exposure time $\Delta t$ in log scale.

\subsection{Mutual relationship between robustness and concentration of signals}

For a deeper insight into the quantitative relationship between optical pre-processing and the immunity of visual inference to detection noise, we explore two scenarios of training networks and the corresponding evaluation methods. First, we investigate the influence of a predefined degree of concentration on the system's resilience against noise during test inference. Second, we reciprocally assess the impact of training noise during the optimization process on the resulting optical network's signal condensation.

\begin{figure}
    \centering
    \includegraphics{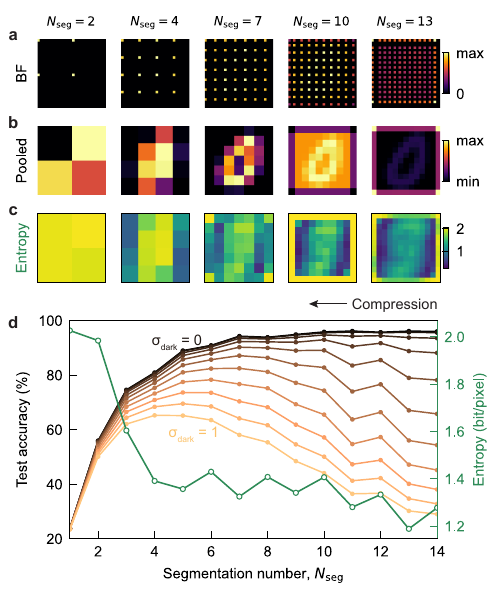}
    \caption{\textbf{Concentration-induced noise robustness.} \textbf{a},\textbf{b}, Output intensity distributions from an input example in class 0 after applying block-wise Fourier operations (\textbf{a}) and then max-pooling (\textbf{b}), with different segmentation numbers $N_\mathrm{seg}=2$ (left) to $13$ (right). \textbf{c}, Shannon entropy distributions given a dataset and the operation with different $N_\mathrm{seg}$. \textbf{d}, MNIST classification accuracies as a function of $N_\mathrm{seg}$ with different test noise levels, from $I_\mathrm{dark}=0$ (black line) to $I_\mathrm{dark}=1$ (orange line). The average entropy per pixel is overlaid.}
    \label{fig3}
\end{figure}

For the first approach, We examine the block-wise Fourier transform of images with various numbers of segmentation, $N_\mathrm{seg}$. Given that the original image consists of $28^2$ pixels, we can consider a uniform segmentation along the width and height of the image with $N_\mathrm{seg} = 1$, 2, 4, 7, and 14, which are all divisors of 28. Otherwise, a non-uniform segmentation is explored as well, for instance, $N_\mathrm{seg}=10$ for dividing 28 into 8 segments of width 3 and 2 segments of width 2 ($28 = 8\times 3 +2\times 2$). Figures 3a and 3b, respectively, depict several intensity distributions $I_\alpha^\mathrm{(out)}$ for block-wise Fourier operation $P_{\mathrm{BF},N_\mathrm{seg}}$ and the corresponding max-pooled images for $N_\mathrm{seg}= 2, 4, 7, 10$, and 13. As $N_\mathrm{seg}$ decreases, notably, the output image becomes more compressive with the dimension reduced to $N_\mathrm{seg}^2$ pixels. Figure 3d validates the noise robustness achieved through compression by presenting the classification accuracy as a function of $N_\mathrm{seg}$ and the test noise power $\sigma_\mathrm{dark}$. As anticipated, more compressive processing with larger $N_\mathrm{seg}$ leads to more robust classification accuracy. This is evidenced in the narrowing gap between the results for zero ($\sigma_\mathrm{dark}=0$, black) and high test noise levels ($\sigma_\mathrm{dark}=1$, yellow). However, it is noted that the ideal accuracy for zero test noise (black line) itself decreases due to the information loss caused by over-compression.

Notably, the way how it is segmented also impacts the overall task performance. Given a dataset, the amount of information in a single focused pixel for each block depends on the size and location of the block. This dependency does not necessarily exhibit an apparent trend with $N_\mathrm{seg}$ but instead fluctuates. To quantify this, we calculate the Shannon entropy \cite{Jaynes1957} for each pixel as a measure of information:
\begin{equation}
    H_\alpha \equiv -\int_{-\infty}^\infty dJ p_\alpha(J)\log_2{p_\alpha(J)},
\end{equation}
where $J_\alpha = \qty[I_\alpha - \mathrm{mean}(I_\alpha)]/\mathrm{Var}(I_\alpha)^{1/2}$ represents the batch-normalized intensity of pixel $\alpha$ over the given validation set, and $p_\alpha$ is the probability distribution function for $J_\alpha$. For example, if a pixel consistently produces a single output intensity regardless of the input class, $H_\alpha =0$. On the contrary, an ideal pixel perfectly classifying into ten different output values depending on the input class has $H_\alpha \sim 3.3$ bits of information. A more compressive, max-pooled pixel per block covers a broader region of input, which tends to include more information with higher entropy. However, pixel-wise entropy fluctuates in practice as indicated by the colour variation in Fig. 3c and the green line in Fig. 3d on average. This fluctuation is likely to impact the total information transferred to the digital network, thereby affecting the overall noise performance, especially for $N_\mathrm{seg}\geq 6$.

\begin{figure}
    \centering
    \includegraphics{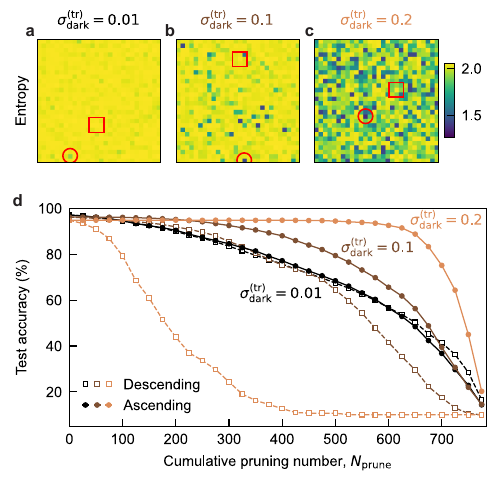}
    \caption{\textbf{Training noise-induced emergence of hub detectors.} \textbf{a}-\textbf{c}, Shannon entropy distributions for trained $\mathrm{U}(28^2)$ operations with the same random initialization but different training noise levels $\sigma_\mathrm{dark}^\mathrm{(tr)}=0.01$ (\textbf{a}), $0.1$ (\textbf{b}), and $0.2$ (\textbf{c}). Red circles and squares indicate the pixels with minimum and maximum entropy of each network, respectively. \textbf{d}, MNIST classification accuracies according to the cumulative pruning of pixels (i.e., enforced zero output to the digital network regardless of input) with ascending (filled circles) or descending (empty square) order of entropy.}
    \label{fig4}
\end{figure}

In the opposite direction, dark noise can induce a general optical linear transformation to be trained in a more compressive manner. In other words, when strong dark noise is applied during training, the output intensity distribution is more likely to be focused on fewer pixels with high SNR based on the select-and-concentrate strategy. Figures 4a-4c illustrate the entropy distribution over pixels trained from the same random initialization $P_\mathrm{R}\in \mathrm{U}(28^2)$ but with different training noise powers $\sigma_\mathrm{dark}^\mathrm{(tr)}$ applied during the optimization process. While almost zero noise (Fig. 4a) results in the equitable optimizations of all pixels in terms of the degree of information contained as represented by mostly flat yellow colours, strong noise (Fig. 4c) leads to differential optimizations over pixels, separating more (yellow) and less (navy) informative pixels.

The concept of noise-induced compression can be proven by ``pruning \cite{Han2015, Yu2023} detections,'' indirectly revealing the contribution of each pixel to the final inference. The pruning of a pixel means to nullify the corresponding detection, by transferring only the pre-calculated batch-mean intensity $\bar{x}_\alpha = \mathrm{mean}[I_\alpha^\mathrm{(out)}]$ instead of the exact detection value $x_\alpha$. Starting from the minimum-entropy pixel, cumulative pruning of pixels in ascending order for the high-noise model (Fig. 4c) does not significantly affect the classification performance until about 600 pixels are eliminated, as depicted by the orange solid line in Fig. 4d. This result implies that the OPU is trained in a way that only around 200 detectors are meaningful. Pruning in the opposite (descending) order beginning with the most important detection, however, results in a rapid accuracy drop (orange dashed line) for the initial 400 cumulative prunings and eventually leads to $\sim10\%$ accuracy which is equivalent to the random guessing. In sharp contrast, the model with almost zero training noise (Fig. 4a) undergoes a more linear-like performance degradation upon the cumulative pruning of detectors both in descending (black dashed line) and ascending (black solid line) orders. These results show the training noise-induced emergence of hub (high entropy) and periphery (low entropy) detectors of differential importance.

\subsection{Practical example: incoherent meta-imaging system}

We have analysed several conceptual results in the discrete model concerning the significance of optical pre-processing in mitigating vulnerability against dark noise. To validate our theoretical approach, we present a practical demonstration through the design of diffractive optical systems, termed the meta-imaging system. This system exhibits superior tolerance to the dark noise compared to conventional imaging devices such as a simple $4f$ system. 

\begin{figure*}
    \centering
    \includegraphics{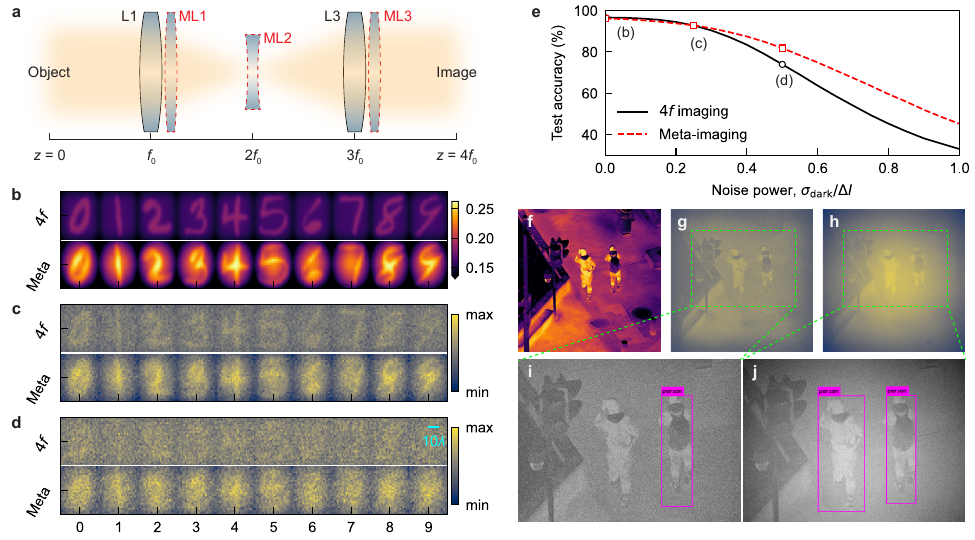}
    \caption{\textbf{Incoherent meta-imaging systems}. \textbf{a}, Illustrations of a conventional 4$f$ system (lenses; L1 and L3) and a meta-imaging system with additional trainable phase masks (metalenses; ML1-3). \textbf{b}-\textbf{d}, Pure images without noise (\textbf{b}) and noisy images with dark noise power $\sigma_\mathrm{dark}=\Delta I/4$ (\textbf{c}) and $\Delta I/2$ (\textbf{d}), obtained by $4f$ (upper) and the optimized meta-imaging (lower) systems for digits 0 to 9. \textbf{e}, MNIST classification accuracies of the conventional (black) and the meta-images (red) as a function of dark noise power. $\lambda$, wavelength; $f_0=300\lambda$ and $\mathrm{NA}\sim0.22$, focal length and numerical aperture of L1 and L3; $\Delta I\sim0.051$, constant for the intensity contrast in conventional images. \textbf{f}-\textbf{j}, Example IR images in reality: a scene of pedestrians (\textbf{f}) from LLVIP dataset \cite{Jia2021}; its modified images based on the pixel-wise intensity ranges for the $4f$ (\textbf{g}) and the optimized meta-imaging (\textbf{h}) systems with the same level of additional Gaussian noises; and the object detection results (magenta boxes, \textbf{i} and \textbf{j}) using the YOLOv3 model\cite{yolov3} for \textbf{g} and \textbf{h}, respectively. Each image is normalized with its minimum and maximum values.} 
    \label{fig5}
\end{figure*}

Meanwhile, it is worth remarking on the difference between coherent and incoherent systems. Let us suppose a spatially incoherent input $E^\mathrm{(in)}_\alpha(t) = [I_\alpha^\mathrm{(in)}]^{1/2} \exp\qty[i\phi_\alpha(t)]$ with a constant intensity $I_\alpha^\mathrm{(in)}$ and a time-varying phase $\phi_\alpha(t)$, extending our discussion to the more realistic passive environment where light usually originates from incoherent sources such as surface emission by the blackbody radiation \cite{Mehta1964}. Given the assumption, the linear field relation $\vb E^\mathrm{(out)} = P \vb E^\mathrm{(in)}$ derives a linear intensity relation\cite{Rahman2023} based on the time-average over a sufficiently long period:
\begin{equation}
    \ev*{|E_\alpha^\mathrm{(out)}|^2}_t = \sum_\beta |P_{\alpha\beta}|^2 |E_\beta^\mathrm{(in)}|^2,
    \label{eq:intensity}
\end{equation}
or, simply $\langle \vb I^\mathrm{(out)} \rangle_t = S \vb I^\mathrm{(in)} $, where $\langle \cdot \rangle_t$ denotes the time average and $S_{\alpha\beta} = |P_{\alpha\beta}|^2$ (See Supplementary Note S1 for derivation and Note S2 for a detailed focusing example). 

As mentioned earlier, optical imaging systems such as the $4f$ system depicted in Fig. 5a are typically linear, which allows us to describe the system through the linear operation between the electric field distributions at input (object, $z=0$) and output (image, $z=4f_0$) planes for a coherent input, or through the linear intensity relation for an incoherent input as well. Especially, when two convex lenses (L1 and L3) of parabolic phase profile $\Phi(x,y)= -\pi(x^2+y^2)/\lambda f_0$ are placed at $z=f_0$ and $3f_0$, where $\lambda$ and $f_0$ are the wavelength and the focal length, respectively, the system operates as an ideal imager. The upper rows of Figs. 5b-5d illustrate the low-contrast incoherent images of MNIST objects without noise (Fig. 5b) and with a weak (Fig. 5c) and strong dark noise (Fig. 5d), calculated using Eq. \eqref{eq:intensity} (see Methods for derivation of $P$). Notably, the coherence length is naturally limited during the numerical calculations as $\ev{E(\vb r_1)E^*(\vb r_2)} = 0$ if $|\vb r_1 - \vb r_2|\geq\sqrt{2}\Delta x$, where $\Delta x = 1.5\lambda$ is the sampling distance chosen for this study.

While the $4f$ system produces a clear image before detection (Fig. 5b), its low-intensity contrast is insufficient to withstand the pronounced detection noise (Fig. 5d). To address this issue with the same strategy of concentrating optical energy into smaller meaningful regions, we introduce additional phase-shift masks, namely metalenses \cite{Khorasaninejad2017}. Positioned at $z=f_0$, $2f_0$, and $3f_0$, three metalenses (ML1-3, Fig. 5a) contribute additional degrees of freedom to the overall classifier, allowing for the optimization of phase profiles to achieve higher classification accuracy in the presence of detection noise (see optimization results in Supplementary Fig. S5). As a result, the trained meta-imaging system generates high-contrast incoherent images by vignetting the less informative area around the four corners while simultaneously highlighting the central part to mitigate the impact of dark noise (bottom row of Fig. 5b). This enhanced intensity contrast certainly improves the machine perception of objects in the presence of strong dark noise, as compared in Fig. 5d. Figure 5e further quantifies the noticeable enhancement in the noise resilience achieved through the metalens-assisted classification (red dashed line), outperforming the conventional $4f$ imaging system (black line).

Our concept remains relevant for inspiring the development of practical machine vision systems, as demonstrated in Figs. 5f-5j. Despite the lack of realistic considerations such as spectral and geometrical parameters in the previous $4f$ and meta-imaging systems and entirely different target purposes as well, their pre-detection modulation behaviours, specifically concentrating optical energy on the central part, can be adapted to a scene \cite{Jia2021} of two pedestrians captured by an IR camera (Fig. 5f). This gives rise to vignetting and Gaussian noise in the images produced by the $4f$ (Fig. 5g) and the meta-imaging (Fig. 5h) systems. Similar to the comparison shown in Fig. 5b, the meta-image in Fig. 5h exhibits a brighter central area, which facilitates machine vision applications in industrial settings. Indeed, employing a pre-trained model (YOLOv3 \cite{yolov3}) allows for the detection of both pedestrians (Fig. 5j) in the noisy meta-image (Fig. 5h), while only one pedestrian (Fig. 5i) is detectable in the conventional image (Fig. 5g) with the same level of Gaussian noise. It is noted that the result shown in Figs. 5f-5j is not optimal but merely a single example showing a clear difference, demanding detailed optimization with a feasible design strategy as future work.

\section{Discussion}
We note that thermal and infrared waves are mostly incoherent for practical uses, such as imaging and vision systems, due to their origin of blackbody radiation. In our prototypical demonstration of noise-robust incoherent meta-imaging in Fig. 5, we have utilized the direct optimization of intensity-intensity linear relation. On the contrary, it is expected to be more rigorous to employ the indirect optimization of electric field-field relation using random phases as discussed in Supplementary Note S1 and Ref. \cite{Rahman2023}, when considering not only the time-average but also the fluctuation of incoherent intensities in a short detection time frame.

To summarize, we have verified the crucial role of optical processing in advance of detection, which concentrates the optical signal power into a smaller region to address the low SNR challenge in noisy systems, such as infrared devices. Through optical computing, the information redundancy in the original distribution of signal power is eliminated until the target performance is not maintained, while the detection power per detector is amplified due to the conservation of total signal energy. Compared to the ideal imaging model where the optical signal is mainly obscured by the severe dark noise, our proposed machine-learned and manually defined optical operations have demonstrated the ability to strategically redistribute optical signals to effectively compete with noise. This outcome underscores the imperative need for harnessing optical computation resources, not only for ultra-fast and energy-efficient bosonic computing but also to navigate noisy environments that cannot be adequately addressed solely through post-detection digital processing.




\bibliography{references}

\begin{thebibliography}{45}%
\makeatletter
\providecommand \@ifxundefined [1]{%
 \@ifx{#1\undefined}
}%
\providecommand \@ifnum [1]{%
 \ifnum #1\expandafter \@firstoftwo
 \else \expandafter \@secondoftwo
 \fi
}%
\providecommand \@ifx [1]{%
 \ifx #1\expandafter \@firstoftwo
 \else \expandafter \@secondoftwo
 \fi
}%
\providecommand \natexlab [1]{#1}%
\providecommand \enquote  [1]{``#1''}%
\providecommand \bibnamefont  [1]{#1}%
\providecommand \bibfnamefont [1]{#1}%
\providecommand \citenamefont [1]{#1}%
\providecommand \href@noop [0]{\@secondoftwo}%
\providecommand \href [0]{\begingroup \@sanitize@url \@href}%
\providecommand \@href[1]{\@@startlink{#1}\@@href}%
\providecommand \@@href[1]{\endgroup#1\@@endlink}%
\providecommand \@sanitize@url [0]{\catcode `\\12\catcode `\$12\catcode `\&12\catcode `\#12\catcode `\^12\catcode `\_12\catcode `\%12\relax}%
\providecommand \@@startlink[1]{}%
\providecommand \@@endlink[0]{}%
\providecommand \url  [0]{\begingroup\@sanitize@url \@url }%
\providecommand \@url [1]{\endgroup\@href {#1}{\urlprefix }}%
\providecommand \urlprefix  [0]{URL }%
\providecommand \Eprint [0]{\href }%
\providecommand \doibase [0]{https://doi.org/}%
\providecommand \selectlanguage [0]{\@gobble}%
\providecommand \bibinfo  [0]{\@secondoftwo}%
\providecommand \bibfield  [0]{\@secondoftwo}%
\providecommand \translation [1]{[#1]}%
\providecommand \BibitemOpen [0]{}%
\providecommand \bibitemStop [0]{}%
\providecommand \bibitemNoStop [0]{.\EOS\space}%
\providecommand \EOS [0]{\spacefactor3000\relax}%
\providecommand \BibitemShut  [1]{\csname bibitem#1\endcsname}%
\let\auto@bib@innerbib\@empty
\bibitem [{\citenamefont {Rogalski}(2010)}]{Rogalski2010}%
  \BibitemOpen
  \bibfield  {author} {\bibinfo {author} {\bibfnamefont {A.}~\bibnamefont {Rogalski}},\ }\href {https://doi.org/10.1201/b10319} {\emph {\bibinfo {title} {Infrared Detectors}}},\ \bibinfo {edition} {2nd}\ ed.\ (\bibinfo  {publisher} {CRC Press},\ \bibinfo {address} {Boca Raton},\ \bibinfo {year} {2010})\BibitemShut {NoStop}%
\bibitem [{\citenamefont {Yang}\ \emph {et~al.}(2021)\citenamefont {Yang}, \citenamefont {Zhang}, \citenamefont {Zhang}, \citenamefont {Wang}, \citenamefont {Feng},\ and\ \citenamefont {Li}}]{Yang2021}%
  \BibitemOpen
  \bibfield  {author} {\bibinfo {author} {\bibfnamefont {J.}~\bibnamefont {Yang}}, \bibinfo {author} {\bibfnamefont {X.}~\bibnamefont {Zhang}}, \bibinfo {author} {\bibfnamefont {X.}~\bibnamefont {Zhang}}, \bibinfo {author} {\bibfnamefont {L.}~\bibnamefont {Wang}}, \bibinfo {author} {\bibfnamefont {W.}~\bibnamefont {Feng}},\ and\ \bibinfo {author} {\bibfnamefont {Q.}~\bibnamefont {Li}},\ }\bibfield  {title} {\bibinfo {title} {Beyond the visible: Bioinspired infrared adaptive materials},\ }\href {https://doi.org/10.1002/adma.202004754} {\bibfield  {journal} {\bibinfo  {journal} {Adv. Mater.}\ }\textbf {\bibinfo {volume} {33}},\ \bibinfo {pages} {2004754} (\bibinfo {year} {2021})}\BibitemShut {NoStop}%
\bibitem [{\citenamefont {Bao}\ \emph {et~al.}(2023)\citenamefont {Bao}, \citenamefont {Wang}, \citenamefont {Sureshbabu}, \citenamefont {Sreekumar}, \citenamefont {Yang}, \citenamefont {Aggarwal}, \citenamefont {Boddeti},\ and\ \citenamefont {Jacob}}]{Bao2023}%
  \BibitemOpen
  \bibfield  {author} {\bibinfo {author} {\bibfnamefont {F.}~\bibnamefont {Bao}}, \bibinfo {author} {\bibfnamefont {X.}~\bibnamefont {Wang}}, \bibinfo {author} {\bibfnamefont {S.~H.}\ \bibnamefont {Sureshbabu}}, \bibinfo {author} {\bibfnamefont {G.}~\bibnamefont {Sreekumar}}, \bibinfo {author} {\bibfnamefont {L.}~\bibnamefont {Yang}}, \bibinfo {author} {\bibfnamefont {V.}~\bibnamefont {Aggarwal}}, \bibinfo {author} {\bibfnamefont {V.~N.}\ \bibnamefont {Boddeti}},\ and\ \bibinfo {author} {\bibfnamefont {Z.}~\bibnamefont {Jacob}},\ }\bibfield  {title} {\bibinfo {title} {Heat-assisted detection and ranging},\ }\href {https://doi.org/10.1038/s41586-023-06174-6} {\bibfield  {journal} {\bibinfo  {journal} {Nature}\ }\textbf {\bibinfo {volume} {619}},\ \bibinfo {pages} {743} (\bibinfo {year} {2023})}\BibitemShut {NoStop}%
\bibitem [{\citenamefont {Marnissi}\ and\ \citenamefont {Fathallah}(2023)}]{Marnissi2023}%
  \BibitemOpen
  \bibfield  {author} {\bibinfo {author} {\bibfnamefont {M.~A.}\ \bibnamefont {Marnissi}}\ and\ \bibinfo {author} {\bibfnamefont {A.}~\bibnamefont {Fathallah}},\ }\bibfield  {title} {\bibinfo {title} {Gan-based vision transformer for high-quality thermal image enhancement},\ }in\ \href@noop {} {\emph {\bibinfo {booktitle} {Proceedings of the IEEE/CVF Conference on Computer Vision and Pattern Recognition (CVPR) Workshops}}}\ (\bibinfo {year} {2023})\ pp.\ \bibinfo {pages} {817--825}\BibitemShut {NoStop}%
\bibitem [{\citenamefont {Raman}\ \emph {et~al.}(2014)\citenamefont {Raman}, \citenamefont {Anoma}, \citenamefont {Zhu}, \citenamefont {Rephaeli},\ and\ \citenamefont {Fan}}]{Raman2014}%
  \BibitemOpen
  \bibfield  {author} {\bibinfo {author} {\bibfnamefont {A.~P.}\ \bibnamefont {Raman}}, \bibinfo {author} {\bibfnamefont {M.~A.}\ \bibnamefont {Anoma}}, \bibinfo {author} {\bibfnamefont {L.}~\bibnamefont {Zhu}}, \bibinfo {author} {\bibfnamefont {E.}~\bibnamefont {Rephaeli}},\ and\ \bibinfo {author} {\bibfnamefont {S.}~\bibnamefont {Fan}},\ }\bibfield  {title} {\bibinfo {title} {Passive radiative cooling below ambient air temperature under direct sunlight},\ }\href {https://doi.org/10.1038/nature13883} {\bibfield  {journal} {\bibinfo  {journal} {Nature}\ }\textbf {\bibinfo {volume} {515}},\ \bibinfo {pages} {540–544} (\bibinfo {year} {2014})}\BibitemShut {NoStop}%
\bibitem [{\citenamefont {Fan}\ and\ \citenamefont {Li}(2022)}]{Fan2022}%
  \BibitemOpen
  \bibfield  {author} {\bibinfo {author} {\bibfnamefont {S.}~\bibnamefont {Fan}}\ and\ \bibinfo {author} {\bibfnamefont {W.}~\bibnamefont {Li}},\ }\bibfield  {title} {\bibinfo {title} {Photonics and thermodynamics concepts in radiative cooling},\ }\href {https://doi.org/10.1038/s41566-021-00921-9} {\bibfield  {journal} {\bibinfo  {journal} {Nat. Photonics}\ }\textbf {\bibinfo {volume} {16}},\ \bibinfo {pages} {182–190} (\bibinfo {year} {2022})}\BibitemShut {NoStop}%
\bibitem [{\citenamefont {Zhao}\ \emph {et~al.}(2023{\natexlab{a}})\citenamefont {Zhao}, \citenamefont {Li}, \citenamefont {Xie}, \citenamefont {Liu}, \citenamefont {Wang}, \citenamefont {Qu}, \citenamefont {Li}, \citenamefont {Liu}, \citenamefont {Brozena}, \citenamefont {Yu}, \citenamefont {Srebric},\ and\ \citenamefont {Hu}}]{xZhao2023}%
  \BibitemOpen
  \bibfield  {author} {\bibinfo {author} {\bibfnamefont {X.}~\bibnamefont {Zhao}}, \bibinfo {author} {\bibfnamefont {T.}~\bibnamefont {Li}}, \bibinfo {author} {\bibfnamefont {H.}~\bibnamefont {Xie}}, \bibinfo {author} {\bibfnamefont {H.}~\bibnamefont {Liu}}, \bibinfo {author} {\bibfnamefont {L.}~\bibnamefont {Wang}}, \bibinfo {author} {\bibfnamefont {Y.}~\bibnamefont {Qu}}, \bibinfo {author} {\bibfnamefont {S.~C.}\ \bibnamefont {Li}}, \bibinfo {author} {\bibfnamefont {S.}~\bibnamefont {Liu}}, \bibinfo {author} {\bibfnamefont {A.~H.}\ \bibnamefont {Brozena}}, \bibinfo {author} {\bibfnamefont {Z.}~\bibnamefont {Yu}}, \bibinfo {author} {\bibfnamefont {J.}~\bibnamefont {Srebric}},\ and\ \bibinfo {author} {\bibfnamefont {L.}~\bibnamefont {Hu}},\ }\bibfield  {title} {\bibinfo {title} {A solution-processed radiative cooling glass},\ }\href {https://doi.org/10.1126/science.adi2224} {\bibfield  {journal} {\bibinfo  {journal} {Science}\ }\textbf {\bibinfo {volume} {382}},\ \bibinfo {pages} {684–691} (\bibinfo {year}
  {2023}{\natexlab{a}})}\BibitemShut {NoStop}%
\bibitem [{\citenamefont {Munday}(2019)}]{Munday2019}%
  \BibitemOpen
  \bibfield  {author} {\bibinfo {author} {\bibfnamefont {J.~N.}\ \bibnamefont {Munday}},\ }\bibfield  {title} {\bibinfo {title} {Tackling climate change through radiative cooling},\ }\href {https://doi.org/10.1016/j.joule.2019.07.010} {\bibfield  {journal} {\bibinfo  {journal} {Joule}\ }\textbf {\bibinfo {volume} {3}},\ \bibinfo {pages} {2057–2060} (\bibinfo {year} {2019})}\BibitemShut {NoStop}%
\bibitem [{\citenamefont {Amenabar}\ \emph {et~al.}(2017)\citenamefont {Amenabar}, \citenamefont {Poly}, \citenamefont {Goikoetxea}, \citenamefont {Nuansing}, \citenamefont {Lasch},\ and\ \citenamefont {Hillenbrand}}]{Amenabar2017}%
  \BibitemOpen
  \bibfield  {author} {\bibinfo {author} {\bibfnamefont {I.}~\bibnamefont {Amenabar}}, \bibinfo {author} {\bibfnamefont {S.}~\bibnamefont {Poly}}, \bibinfo {author} {\bibfnamefont {M.}~\bibnamefont {Goikoetxea}}, \bibinfo {author} {\bibfnamefont {W.}~\bibnamefont {Nuansing}}, \bibinfo {author} {\bibfnamefont {P.}~\bibnamefont {Lasch}},\ and\ \bibinfo {author} {\bibfnamefont {R.}~\bibnamefont {Hillenbrand}},\ }\bibfield  {title} {\bibinfo {title} {Hyperspectral infrared nanoimaging of organic samples based on fourier transform infrared nanospectroscopy},\ }\href {https://doi.org/10.1038/ncomms14402} {\bibfield  {journal} {\bibinfo  {journal} {Nat. Commun.}\ }\textbf {\bibinfo {volume} {8}},\ \bibinfo {pages} {14402} (\bibinfo {year} {2017})}\BibitemShut {NoStop}%
\bibitem [{\citenamefont {Wang}\ \emph {et~al.}(2019)\citenamefont {Wang}, \citenamefont {Yi}, \citenamefont {Chen}, \citenamefont {Zhou}, \citenamefont {Luk}, \citenamefont {James}, \citenamefont {Nogan}, \citenamefont {Ross}, \citenamefont {Joe}, \citenamefont {Shahsafi}, \citenamefont {Wang}, \citenamefont {Kats},\ and\ \citenamefont {Yu}}]{Wang2019}%
  \BibitemOpen
  \bibfield  {author} {\bibinfo {author} {\bibfnamefont {Z.}~\bibnamefont {Wang}}, \bibinfo {author} {\bibfnamefont {S.}~\bibnamefont {Yi}}, \bibinfo {author} {\bibfnamefont {A.}~\bibnamefont {Chen}}, \bibinfo {author} {\bibfnamefont {M.}~\bibnamefont {Zhou}}, \bibinfo {author} {\bibfnamefont {T.~S.}\ \bibnamefont {Luk}}, \bibinfo {author} {\bibfnamefont {A.}~\bibnamefont {James}}, \bibinfo {author} {\bibfnamefont {J.}~\bibnamefont {Nogan}}, \bibinfo {author} {\bibfnamefont {W.}~\bibnamefont {Ross}}, \bibinfo {author} {\bibfnamefont {G.}~\bibnamefont {Joe}}, \bibinfo {author} {\bibfnamefont {A.}~\bibnamefont {Shahsafi}}, \bibinfo {author} {\bibfnamefont {K.~X.}\ \bibnamefont {Wang}}, \bibinfo {author} {\bibfnamefont {M.~A.}\ \bibnamefont {Kats}},\ and\ \bibinfo {author} {\bibfnamefont {Z.}~\bibnamefont {Yu}},\ }\bibfield  {title} {\bibinfo {title} {Single-shot on-chip spectral sensors based on photonic crystal slabs},\ }\href {https://doi.org/10.1038/s41467-019-08994-5} {\bibfield  {journal} {\bibinfo
  {journal} {Nat. Commun.}\ }\textbf {\bibinfo {volume} {10}},\ \bibinfo {pages} {1020} (\bibinfo {year} {2019})}\BibitemShut {NoStop}%
\bibitem [{\citenamefont {Zhao}\ \emph {et~al.}(2023{\natexlab{b}})\citenamefont {Zhao}, \citenamefont {Kusama}, \citenamefont {Furutani}, \citenamefont {Huang}, \citenamefont {Luo},\ and\ \citenamefont {Fuji}}]{Zhao2023}%
  \BibitemOpen
  \bibfield  {author} {\bibinfo {author} {\bibfnamefont {Y.}~\bibnamefont {Zhao}}, \bibinfo {author} {\bibfnamefont {S.}~\bibnamefont {Kusama}}, \bibinfo {author} {\bibfnamefont {Y.}~\bibnamefont {Furutani}}, \bibinfo {author} {\bibfnamefont {W.-H.}\ \bibnamefont {Huang}}, \bibinfo {author} {\bibfnamefont {C.-W.}\ \bibnamefont {Luo}},\ and\ \bibinfo {author} {\bibfnamefont {T.}~\bibnamefont {Fuji}},\ }\bibfield  {title} {\bibinfo {title} {High-speed scanless entire bandwidth mid-infrared chemical imaging},\ }\href {https://doi.org/10.1038/s41467-023-39628-6} {\bibfield  {journal} {\bibinfo  {journal} {Nat. Commun.}\ }\textbf {\bibinfo {volume} {14}},\ \bibinfo {pages} {3929} (\bibinfo {year} {2023}{\natexlab{b}})}\BibitemShut {NoStop}%
\bibitem [{\citenamefont {Wang}\ \emph {et~al.}(2024)\citenamefont {Wang}, \citenamefont {Yang}, \citenamefont {Bao}, \citenamefont {Sentz},\ and\ \citenamefont {Jacob}}]{xWang2023}%
  \BibitemOpen
  \bibfield  {author} {\bibinfo {author} {\bibfnamefont {X.}~\bibnamefont {Wang}}, \bibinfo {author} {\bibfnamefont {Z.}~\bibnamefont {Yang}}, \bibinfo {author} {\bibfnamefont {F.}~\bibnamefont {Bao}}, \bibinfo {author} {\bibfnamefont {T.}~\bibnamefont {Sentz}},\ and\ \bibinfo {author} {\bibfnamefont {Z.}~\bibnamefont {Jacob}},\ }\bibfield  {title} {\bibinfo {title} {Spinning metasurface stack for spectro-polarimetric thermal imaging},\ }\href {https://doi.org/10.1364/OPTICA.506813} {\bibfield  {journal} {\bibinfo  {journal} {Optica}\ }\textbf {\bibinfo {volume} {11}},\ \bibinfo {pages} {73} (\bibinfo {year} {2024})}\BibitemShut {NoStop}%
\bibitem [{\citenamefont {Gurton}\ \emph {et~al.}(2014)\citenamefont {Gurton}, \citenamefont {Yuffa},\ and\ \citenamefont {Videen}}]{Gurton2014}%
  \BibitemOpen
  \bibfield  {author} {\bibinfo {author} {\bibfnamefont {K.~P.}\ \bibnamefont {Gurton}}, \bibinfo {author} {\bibfnamefont {A.~J.}\ \bibnamefont {Yuffa}},\ and\ \bibinfo {author} {\bibfnamefont {G.~W.}\ \bibnamefont {Videen}},\ }\bibfield  {title} {\bibinfo {title} {Enhanced facial recognition for thermal imagery using polarimetric imaging},\ }\href {https://doi.org/10.1364/ol.39.003857} {\bibfield  {journal} {\bibinfo  {journal} {Opt. Lett.}\ }\textbf {\bibinfo {volume} {39}},\ \bibinfo {pages} {3857} (\bibinfo {year} {2014})}\BibitemShut {NoStop}%
\bibitem [{\citenamefont {Yuffa}\ \emph {et~al.}(2014)\citenamefont {Yuffa}, \citenamefont {Gurton},\ and\ \citenamefont {Videen}}]{Yuffa2014}%
  \BibitemOpen
  \bibfield  {author} {\bibinfo {author} {\bibfnamefont {A.~J.}\ \bibnamefont {Yuffa}}, \bibinfo {author} {\bibfnamefont {K.~P.}\ \bibnamefont {Gurton}},\ and\ \bibinfo {author} {\bibfnamefont {G.}~\bibnamefont {Videen}},\ }\bibfield  {title} {\bibinfo {title} {Three-dimensional facial recognition using passive long-wavelength infrared polarimetric imaging},\ }\href {https://doi.org/10.1364/ao.53.008514} {\bibfield  {journal} {\bibinfo  {journal} {Appl. Opt.}\ }\textbf {\bibinfo {volume} {53}},\ \bibinfo {pages} {8514} (\bibinfo {year} {2014})}\BibitemShut {NoStop}%
\bibitem [{\citenamefont {Deng}\ \emph {et~al.}(2022)\citenamefont {Deng}, \citenamefont {Dai}, \citenamefont {Wang}, \citenamefont {You}, \citenamefont {Chen}, \citenamefont {Han}, \citenamefont {Han}, \citenamefont {Wang}, \citenamefont {Ye}, \citenamefont {Zhu}, \citenamefont {Cui}, \citenamefont {Wang},\ and\ \citenamefont {Zhang}}]{Deng2022}%
  \BibitemOpen
  \bibfield  {author} {\bibinfo {author} {\bibfnamefont {W.}~\bibnamefont {Deng}}, \bibinfo {author} {\bibfnamefont {M.}~\bibnamefont {Dai}}, \bibinfo {author} {\bibfnamefont {C.}~\bibnamefont {Wang}}, \bibinfo {author} {\bibfnamefont {C.}~\bibnamefont {You}}, \bibinfo {author} {\bibfnamefont {W.}~\bibnamefont {Chen}}, \bibinfo {author} {\bibfnamefont {S.}~\bibnamefont {Han}}, \bibinfo {author} {\bibfnamefont {J.}~\bibnamefont {Han}}, \bibinfo {author} {\bibfnamefont {F.}~\bibnamefont {Wang}}, \bibinfo {author} {\bibfnamefont {M.}~\bibnamefont {Ye}}, \bibinfo {author} {\bibfnamefont {S.}~\bibnamefont {Zhu}}, \bibinfo {author} {\bibfnamefont {J.}~\bibnamefont {Cui}}, \bibinfo {author} {\bibfnamefont {Q.~J.}\ \bibnamefont {Wang}},\ and\ \bibinfo {author} {\bibfnamefont {Y.}~\bibnamefont {Zhang}},\ }\bibfield  {title} {\bibinfo {title} {Switchable unipolar‐barrier van der waals heterostructures with natural anisotropy for full linear polarimetry detection},\ }\href {https://doi.org/10.1002/adma.202203766}
  {\bibfield  {journal} {\bibinfo  {journal} {Adv. Mater.}\ }\textbf {\bibinfo {volume} {34}},\ \bibinfo {pages} {2203766} (\bibinfo {year} {2022})}\BibitemShut {NoStop}%
\bibitem [{\citenamefont {Szeliski}(2022)}]{Szeliski2022}%
  \BibitemOpen
  \bibfield  {author} {\bibinfo {author} {\bibfnamefont {R.}~\bibnamefont {Szeliski}},\ }\href {https://doi.org/10.1007/978-3-030-34372-9} {\emph {\bibinfo {title} {Computer Vision: Algorithms and Applications}}},\ \bibinfo {edition} {2nd}\ ed.\ (\bibinfo  {publisher} {Springer International Publishing},\ \bibinfo {year} {2022})\BibitemShut {NoStop}%
\bibitem [{\citenamefont {Lin}\ \emph {et~al.}(2018)\citenamefont {Lin}, \citenamefont {Rivenson}, \citenamefont {Yardimci}, \citenamefont {Veli}, \citenamefont {Luo}, \citenamefont {Jarrahi},\ and\ \citenamefont {Ozcan}}]{Lin2018}%
  \BibitemOpen
  \bibfield  {author} {\bibinfo {author} {\bibfnamefont {X.}~\bibnamefont {Lin}}, \bibinfo {author} {\bibfnamefont {Y.}~\bibnamefont {Rivenson}}, \bibinfo {author} {\bibfnamefont {N.~T.}\ \bibnamefont {Yardimci}}, \bibinfo {author} {\bibfnamefont {M.}~\bibnamefont {Veli}}, \bibinfo {author} {\bibfnamefont {Y.}~\bibnamefont {Luo}}, \bibinfo {author} {\bibfnamefont {M.}~\bibnamefont {Jarrahi}},\ and\ \bibinfo {author} {\bibfnamefont {A.}~\bibnamefont {Ozcan}},\ }\bibfield  {title} {\bibinfo {title} {All-optical machine learning using diffractive deep neural networks},\ }\href {https://doi.org/10.1126/science.aat8084} {\bibfield  {journal} {\bibinfo  {journal} {Science}\ }\textbf {\bibinfo {volume} {361}},\ \bibinfo {pages} {1004} (\bibinfo {year} {2018})}\BibitemShut {NoStop}%
\bibitem [{\citenamefont {Wu}\ \emph {et~al.}(2019)\citenamefont {Wu}, \citenamefont {Zhou}, \citenamefont {Khoram}, \citenamefont {Liu},\ and\ \citenamefont {Yu}}]{Wu2019}%
  \BibitemOpen
  \bibfield  {author} {\bibinfo {author} {\bibfnamefont {Z.}~\bibnamefont {Wu}}, \bibinfo {author} {\bibfnamefont {M.}~\bibnamefont {Zhou}}, \bibinfo {author} {\bibfnamefont {E.}~\bibnamefont {Khoram}}, \bibinfo {author} {\bibfnamefont {B.}~\bibnamefont {Liu}},\ and\ \bibinfo {author} {\bibfnamefont {Z.}~\bibnamefont {Yu}},\ }\bibfield  {title} {\bibinfo {title} {Neuromorphic metasurface},\ }\href {https://doi.org/10.1364/prj.8.000046} {\bibfield  {journal} {\bibinfo  {journal} {Photon. Res.}\ }\textbf {\bibinfo {volume} {8}},\ \bibinfo {pages} {46} (\bibinfo {year} {2019})}\BibitemShut {NoStop}%
\bibitem [{\citenamefont {Wu}\ and\ \citenamefont {Yu}(2021)}]{Wu2021}%
  \BibitemOpen
  \bibfield  {author} {\bibinfo {author} {\bibfnamefont {Z.}~\bibnamefont {Wu}}\ and\ \bibinfo {author} {\bibfnamefont {Z.}~\bibnamefont {Yu}},\ }\bibfield  {title} {\bibinfo {title} {Small object recognition with trainable lens},\ }\href {https://doi.org/10.1063/5.0054117} {\bibfield  {journal} {\bibinfo  {journal} {APL Photonics}\ }\textbf {\bibinfo {volume} {6}},\ \bibinfo {pages} {071301} (\bibinfo {year} {2021})}\BibitemShut {NoStop}%
\bibitem [{\citenamefont {Bernstein}\ \emph {et~al.}(2023)\citenamefont {Bernstein}, \citenamefont {Sludds}, \citenamefont {Panuski}, \citenamefont {Trajtenberg-Mills}, \citenamefont {Hamerly},\ and\ \citenamefont {Englund}}]{Bernstein2023}%
  \BibitemOpen
  \bibfield  {author} {\bibinfo {author} {\bibfnamefont {L.}~\bibnamefont {Bernstein}}, \bibinfo {author} {\bibfnamefont {A.}~\bibnamefont {Sludds}}, \bibinfo {author} {\bibfnamefont {C.}~\bibnamefont {Panuski}}, \bibinfo {author} {\bibfnamefont {S.}~\bibnamefont {Trajtenberg-Mills}}, \bibinfo {author} {\bibfnamefont {R.}~\bibnamefont {Hamerly}},\ and\ \bibinfo {author} {\bibfnamefont {D.}~\bibnamefont {Englund}},\ }\bibfield  {title} {\bibinfo {title} {Single-shot optical neural network},\ }\href {https://doi.org/10.1126/sciadv.adg7904} {\bibfield  {journal} {\bibinfo  {journal} {Sci. Adv.}\ }\textbf {\bibinfo {volume} {9}},\ \bibinfo {pages} {eadg7904} (\bibinfo {year} {2023})}\BibitemShut {NoStop}%
\bibitem [{\citenamefont {Chen}\ \emph {et~al.}(2023{\natexlab{a}})\citenamefont {Chen}, \citenamefont {Zhou}, \citenamefont {Wu}, \citenamefont {Qiao}, \citenamefont {Lin}, \citenamefont {Fang},\ and\ \citenamefont {Dai}}]{yChen2023}%
  \BibitemOpen
  \bibfield  {author} {\bibinfo {author} {\bibfnamefont {Y.}~\bibnamefont {Chen}}, \bibinfo {author} {\bibfnamefont {T.}~\bibnamefont {Zhou}}, \bibinfo {author} {\bibfnamefont {J.}~\bibnamefont {Wu}}, \bibinfo {author} {\bibfnamefont {H.}~\bibnamefont {Qiao}}, \bibinfo {author} {\bibfnamefont {X.}~\bibnamefont {Lin}}, \bibinfo {author} {\bibfnamefont {L.}~\bibnamefont {Fang}},\ and\ \bibinfo {author} {\bibfnamefont {Q.}~\bibnamefont {Dai}},\ }\bibfield  {title} {\bibinfo {title} {Photonic unsupervised learning variational autoencoder for high-throughput and low-latency image transmission},\ }\href {https://doi.org/10.1126/sciadv.adf8437} {\bibfield  {journal} {\bibinfo  {journal} {Sci. Adv.}\ }\textbf {\bibinfo {volume} {9}},\ \bibinfo {pages} {eadf8437} (\bibinfo {year} {2023}{\natexlab{a}})}\BibitemShut {NoStop}%
\bibitem [{\citenamefont {Fu}\ \emph {et~al.}(2023)\citenamefont {Fu}, \citenamefont {Zang}, \citenamefont {Huang}, \citenamefont {Du}, \citenamefont {Huang}, \citenamefont {Hu}, \citenamefont {Chen}, \citenamefont {Yang},\ and\ \citenamefont {Chen}}]{Fu2023}%
  \BibitemOpen
  \bibfield  {author} {\bibinfo {author} {\bibfnamefont {T.}~\bibnamefont {Fu}}, \bibinfo {author} {\bibfnamefont {Y.}~\bibnamefont {Zang}}, \bibinfo {author} {\bibfnamefont {Y.}~\bibnamefont {Huang}}, \bibinfo {author} {\bibfnamefont {Z.}~\bibnamefont {Du}}, \bibinfo {author} {\bibfnamefont {H.}~\bibnamefont {Huang}}, \bibinfo {author} {\bibfnamefont {C.}~\bibnamefont {Hu}}, \bibinfo {author} {\bibfnamefont {M.}~\bibnamefont {Chen}}, \bibinfo {author} {\bibfnamefont {S.}~\bibnamefont {Yang}},\ and\ \bibinfo {author} {\bibfnamefont {H.}~\bibnamefont {Chen}},\ }\bibfield  {title} {\bibinfo {title} {Photonic machine learning with on-chip diffractive optics},\ }\href {https://doi.org/10.1038/s41467-022-35772-7} {\bibfield  {journal} {\bibinfo  {journal} {Nat. Commun.}\ }\textbf {\bibinfo {volume} {14}},\ \bibinfo {pages} {70} (\bibinfo {year} {2023})}\BibitemShut {NoStop}%
\bibitem [{\citenamefont {Wang}\ \emph {et~al.}(2023)\citenamefont {Wang}, \citenamefont {Sohoni}, \citenamefont {Wright}, \citenamefont {Stein}, \citenamefont {Ma}, \citenamefont {Onodera}, \citenamefont {Anderson},\ and\ \citenamefont {McMahon}}]{TWang2023}%
  \BibitemOpen
  \bibfield  {author} {\bibinfo {author} {\bibfnamefont {T.}~\bibnamefont {Wang}}, \bibinfo {author} {\bibfnamefont {M.~M.}\ \bibnamefont {Sohoni}}, \bibinfo {author} {\bibfnamefont {L.~G.}\ \bibnamefont {Wright}}, \bibinfo {author} {\bibfnamefont {M.~M.}\ \bibnamefont {Stein}}, \bibinfo {author} {\bibfnamefont {S.-Y.}\ \bibnamefont {Ma}}, \bibinfo {author} {\bibfnamefont {T.}~\bibnamefont {Onodera}}, \bibinfo {author} {\bibfnamefont {M.~G.}\ \bibnamefont {Anderson}},\ and\ \bibinfo {author} {\bibfnamefont {P.~L.}\ \bibnamefont {McMahon}},\ }\bibfield  {title} {\bibinfo {title} {Image sensing with multilayer nonlinear optical neural networks},\ }\href {https://doi.org/10.1038/s41566-023-01170-8} {\bibfield  {journal} {\bibinfo  {journal} {Nat. Photonics}\ }\textbf {\bibinfo {volume} {17}},\ \bibinfo {pages} {408} (\bibinfo {year} {2023})}\BibitemShut {NoStop}%
\bibitem [{\citenamefont {Zheng}\ \emph {et~al.}(2023)\citenamefont {Zheng}, \citenamefont {Duan}, \citenamefont {Chen}, \citenamefont {Yang}, \citenamefont {Gao}, \citenamefont {Zhang}, \citenamefont {Xiong},\ and\ \citenamefont {Lin}}]{Zheng2023}%
  \BibitemOpen
  \bibfield  {author} {\bibinfo {author} {\bibfnamefont {Z.}~\bibnamefont {Zheng}}, \bibinfo {author} {\bibfnamefont {Z.}~\bibnamefont {Duan}}, \bibinfo {author} {\bibfnamefont {H.}~\bibnamefont {Chen}}, \bibinfo {author} {\bibfnamefont {R.}~\bibnamefont {Yang}}, \bibinfo {author} {\bibfnamefont {S.}~\bibnamefont {Gao}}, \bibinfo {author} {\bibfnamefont {H.}~\bibnamefont {Zhang}}, \bibinfo {author} {\bibfnamefont {H.}~\bibnamefont {Xiong}},\ and\ \bibinfo {author} {\bibfnamefont {X.}~\bibnamefont {Lin}},\ }\bibfield  {title} {\bibinfo {title} {Dual adaptive training of photonic neural networks},\ }\href {https://doi.org/10.1038/s42256-023-00723-4} {\bibfield  {journal} {\bibinfo  {journal} {Nat. Mach. Intell.}\ }\textbf {\bibinfo {volume} {5}},\ \bibinfo {pages} {1119} (\bibinfo {year} {2023})}\BibitemShut {NoStop}%
\bibitem [{\citenamefont {Shen}\ \emph {et~al.}(2017)\citenamefont {Shen}, \citenamefont {Harris}, \citenamefont {Skirlo}, \citenamefont {Prabhu}, \citenamefont {Baehr-Jones}, \citenamefont {Hochberg}, \citenamefont {Sun}, \citenamefont {Zhao}, \citenamefont {Larochelle}, \citenamefont {Englund},\ and\ \citenamefont {Solja{\v{c}}i{\'{c}}}}]{Shen2017}%
  \BibitemOpen
  \bibfield  {author} {\bibinfo {author} {\bibfnamefont {Y.}~\bibnamefont {Shen}}, \bibinfo {author} {\bibfnamefont {N.~C.}\ \bibnamefont {Harris}}, \bibinfo {author} {\bibfnamefont {S.}~\bibnamefont {Skirlo}}, \bibinfo {author} {\bibfnamefont {M.}~\bibnamefont {Prabhu}}, \bibinfo {author} {\bibfnamefont {T.}~\bibnamefont {Baehr-Jones}}, \bibinfo {author} {\bibfnamefont {M.}~\bibnamefont {Hochberg}}, \bibinfo {author} {\bibfnamefont {X.}~\bibnamefont {Sun}}, \bibinfo {author} {\bibfnamefont {S.}~\bibnamefont {Zhao}}, \bibinfo {author} {\bibfnamefont {H.}~\bibnamefont {Larochelle}}, \bibinfo {author} {\bibfnamefont {D.}~\bibnamefont {Englund}},\ and\ \bibinfo {author} {\bibfnamefont {M.}~\bibnamefont {Solja{\v{c}}i{\'{c}}}},\ }\bibfield  {title} {\bibinfo {title} {Deep learning with coherent nanophotonic circuits},\ }\href {https://doi.org/10.1038/nphoton.2017.93} {\bibfield  {journal} {\bibinfo  {journal} {Nat. Photonics}\ }\textbf {\bibinfo {volume} {11}},\ \bibinfo {pages} {441} (\bibinfo {year}
  {2017})}\BibitemShut {NoStop}%
\bibitem [{\citenamefont {Hamerly}\ \emph {et~al.}(2019)\citenamefont {Hamerly}, \citenamefont {Bernstein}, \citenamefont {Sludds}, \citenamefont {Solja\ifmmode \check{c}\else \v{c}\fi{}i\ifmmode~\acute{c}\else \'{c}\fi{}},\ and\ \citenamefont {Englund}}]{Hamerly2019}%
  \BibitemOpen
  \bibfield  {author} {\bibinfo {author} {\bibfnamefont {R.}~\bibnamefont {Hamerly}}, \bibinfo {author} {\bibfnamefont {L.}~\bibnamefont {Bernstein}}, \bibinfo {author} {\bibfnamefont {A.}~\bibnamefont {Sludds}}, \bibinfo {author} {\bibfnamefont {M.}~\bibnamefont {Solja\ifmmode \check{c}\else \v{c}\fi{}i\ifmmode~\acute{c}\else \'{c}\fi{}}},\ and\ \bibinfo {author} {\bibfnamefont {D.}~\bibnamefont {Englund}},\ }\bibfield  {title} {\bibinfo {title} {Large-scale optical neural networks based on photoelectric multiplication},\ }\href {https://doi.org/10.1103/PhysRevX.9.021032} {\bibfield  {journal} {\bibinfo  {journal} {Phys. Rev. X}\ }\textbf {\bibinfo {volume} {9}},\ \bibinfo {pages} {021032} (\bibinfo {year} {2019})}\BibitemShut {NoStop}%
\bibitem [{\citenamefont {Mourgias-Alexandris}\ \emph {et~al.}(2022)\citenamefont {Mourgias-Alexandris}, \citenamefont {Moralis-Pegios}, \citenamefont {Tsakyridis}, \citenamefont {Simos}, \citenamefont {Dabos}, \citenamefont {Totovic}, \citenamefont {Passalis}, \citenamefont {Kirtas}, \citenamefont {Rutirawut}, \citenamefont {Gardes}, \citenamefont {Tefas},\ and\ \citenamefont {Pleros}}]{MourgiasAlexandris2022}%
  \BibitemOpen
  \bibfield  {author} {\bibinfo {author} {\bibfnamefont {G.}~\bibnamefont {Mourgias-Alexandris}}, \bibinfo {author} {\bibfnamefont {M.}~\bibnamefont {Moralis-Pegios}}, \bibinfo {author} {\bibfnamefont {A.}~\bibnamefont {Tsakyridis}}, \bibinfo {author} {\bibfnamefont {S.}~\bibnamefont {Simos}}, \bibinfo {author} {\bibfnamefont {G.}~\bibnamefont {Dabos}}, \bibinfo {author} {\bibfnamefont {A.}~\bibnamefont {Totovic}}, \bibinfo {author} {\bibfnamefont {N.}~\bibnamefont {Passalis}}, \bibinfo {author} {\bibfnamefont {M.}~\bibnamefont {Kirtas}}, \bibinfo {author} {\bibfnamefont {T.}~\bibnamefont {Rutirawut}}, \bibinfo {author} {\bibfnamefont {F.~Y.}\ \bibnamefont {Gardes}}, \bibinfo {author} {\bibfnamefont {A.}~\bibnamefont {Tefas}},\ and\ \bibinfo {author} {\bibfnamefont {N.}~\bibnamefont {Pleros}},\ }\bibfield  {title} {\bibinfo {title} {Noise-resilient and high-speed deep learning with coherent silicon photonics},\ }\href {https://doi.org/10.1038/s41467-022-33259-z} {\bibfield  {journal} {\bibinfo  {journal}
  {Nat. Commun.}\ }\textbf {\bibinfo {volume} {13}},\ \bibinfo {pages} {5572} (\bibinfo {year} {2022})}\BibitemShut {NoStop}%
\bibitem [{\citenamefont {Chen}\ \emph {et~al.}(2023{\natexlab{b}})\citenamefont {Chen}, \citenamefont {Sludds}, \citenamefont {Davis}, \citenamefont {Christen}, \citenamefont {Bernstein}, \citenamefont {Ateshian}, \citenamefont {Heuser}, \citenamefont {Heermeier}, \citenamefont {Lott}, \citenamefont {Reitzenstein}, \citenamefont {Hamerly},\ and\ \citenamefont {Englund}}]{ZChen2023}%
  \BibitemOpen
  \bibfield  {author} {\bibinfo {author} {\bibfnamefont {Z.}~\bibnamefont {Chen}}, \bibinfo {author} {\bibfnamefont {A.}~\bibnamefont {Sludds}}, \bibinfo {author} {\bibfnamefont {R.}~\bibnamefont {Davis}}, \bibinfo {author} {\bibfnamefont {I.}~\bibnamefont {Christen}}, \bibinfo {author} {\bibfnamefont {L.}~\bibnamefont {Bernstein}}, \bibinfo {author} {\bibfnamefont {L.}~\bibnamefont {Ateshian}}, \bibinfo {author} {\bibfnamefont {T.}~\bibnamefont {Heuser}}, \bibinfo {author} {\bibfnamefont {N.}~\bibnamefont {Heermeier}}, \bibinfo {author} {\bibfnamefont {J.~A.}\ \bibnamefont {Lott}}, \bibinfo {author} {\bibfnamefont {S.}~\bibnamefont {Reitzenstein}}, \bibinfo {author} {\bibfnamefont {R.}~\bibnamefont {Hamerly}},\ and\ \bibinfo {author} {\bibfnamefont {D.}~\bibnamefont {Englund}},\ }\bibfield  {title} {\bibinfo {title} {Deep learning with coherent vcsel neural networks},\ }\href {https://doi.org/10.1038/s41566-023-01233-w} {\bibfield  {journal} {\bibinfo  {journal} {Nat. Photonics}\ }\textbf {\bibinfo {volume}
  {17}},\ \bibinfo {pages} {723–730} (\bibinfo {year} {2023}{\natexlab{b}})}\BibitemShut {NoStop}%
\bibitem [{\citenamefont {Yu}\ and\ \citenamefont {Park}(2023)}]{Yu2023}%
  \BibitemOpen
  \bibfield  {author} {\bibinfo {author} {\bibfnamefont {S.}~\bibnamefont {Yu}}\ and\ \bibinfo {author} {\bibfnamefont {N.}~\bibnamefont {Park}},\ }\bibfield  {title} {\bibinfo {title} {Heavy tails and pruning in programmable photonic circuits for universal unitaries},\ }\href {https://doi.org/10.1038/s41467-023-37611-9} {\bibfield  {journal} {\bibinfo  {journal} {Nat. Commun.}\ }\textbf {\bibinfo {volume} {14}},\ \bibinfo {pages} {1853} (\bibinfo {year} {2023})}\BibitemShut {NoStop}%
\bibitem [{\citenamefont {Colthup}\ \emph {et~al.}(1990)\citenamefont {Colthup}, \citenamefont {Daly},\ and\ \citenamefont {Wiberley}}]{Colthup1990}%
  \BibitemOpen
  \bibfield  {author} {\bibinfo {author} {\bibfnamefont {N.~B.}\ \bibnamefont {Colthup}}, \bibinfo {author} {\bibfnamefont {L.~H.}\ \bibnamefont {Daly}},\ and\ \bibinfo {author} {\bibfnamefont {S.~E.}\ \bibnamefont {Wiberley}},\ }\bibfield  {title} {\bibinfo {title} {{IR} experimental considerations},\ }in\ \href {https://doi.org/https://doi.org/10.1016/B978-0-08-091740-5.50005-3} {\emph {\bibinfo {booktitle} {Introduction to Infrared and Raman Spectroscopy}}}\ (\bibinfo  {publisher} {Academic Press},\ \bibinfo {address} {San Diego},\ \bibinfo {year} {1990})\ \bibinfo {edition} {3rd}\ ed.,\ pp.\ \bibinfo {pages} {75--107}\BibitemShut {NoStop}%
\bibitem [{\citenamefont {Cohen}\ \emph {et~al.}(2017)\citenamefont {Cohen}, \citenamefont {Afshar}, \citenamefont {Tapson},\ and\ \citenamefont {van Schaik}}]{Cohen2017}%
  \BibitemOpen
  \bibfield  {author} {\bibinfo {author} {\bibfnamefont {G.}~\bibnamefont {Cohen}}, \bibinfo {author} {\bibfnamefont {S.}~\bibnamefont {Afshar}}, \bibinfo {author} {\bibfnamefont {J.}~\bibnamefont {Tapson}},\ and\ \bibinfo {author} {\bibfnamefont {A.}~\bibnamefont {van Schaik}},\ }\href {https://doi.org/10.48550/ARXIV.1702.05373} {\bibinfo {title} {Emnist: an extension of mnist to handwritten letters}} (\bibinfo {year} {2017})\BibitemShut {NoStop}%
\bibitem [{\citenamefont {Carolan}\ \emph {et~al.}(2015)\citenamefont {Carolan}, \citenamefont {Harrold}, \citenamefont {Sparrow}, \citenamefont {Martín-López}, \citenamefont {Russell}, \citenamefont {Silverstone}, \citenamefont {Shadbolt}, \citenamefont {Matsuda}, \citenamefont {Oguma}, \citenamefont {Itoh}, \citenamefont {Marshall}, \citenamefont {Thompson}, \citenamefont {Matthews}, \citenamefont {Hashimoto}, \citenamefont {O’Brien},\ and\ \citenamefont {Laing}}]{Carolan2015}%
  \BibitemOpen
  \bibfield  {author} {\bibinfo {author} {\bibfnamefont {J.}~\bibnamefont {Carolan}}, \bibinfo {author} {\bibfnamefont {C.}~\bibnamefont {Harrold}}, \bibinfo {author} {\bibfnamefont {C.}~\bibnamefont {Sparrow}}, \bibinfo {author} {\bibfnamefont {E.}~\bibnamefont {Martín-López}}, \bibinfo {author} {\bibfnamefont {N.~J.}\ \bibnamefont {Russell}}, \bibinfo {author} {\bibfnamefont {J.~W.}\ \bibnamefont {Silverstone}}, \bibinfo {author} {\bibfnamefont {P.~J.}\ \bibnamefont {Shadbolt}}, \bibinfo {author} {\bibfnamefont {N.}~\bibnamefont {Matsuda}}, \bibinfo {author} {\bibfnamefont {M.}~\bibnamefont {Oguma}}, \bibinfo {author} {\bibfnamefont {M.}~\bibnamefont {Itoh}}, \bibinfo {author} {\bibfnamefont {G.~D.}\ \bibnamefont {Marshall}}, \bibinfo {author} {\bibfnamefont {M.~G.}\ \bibnamefont {Thompson}}, \bibinfo {author} {\bibfnamefont {J.~C.~F.}\ \bibnamefont {Matthews}}, \bibinfo {author} {\bibfnamefont {T.}~\bibnamefont {Hashimoto}}, \bibinfo {author} {\bibfnamefont {J.~L.}\ \bibnamefont {O’Brien}},\ and\
  \bibinfo {author} {\bibfnamefont {A.}~\bibnamefont {Laing}},\ }\bibfield  {title} {\bibinfo {title} {Universal linear optics},\ }\href {https://doi.org/10.1126/science.aab3642} {\bibfield  {journal} {\bibinfo  {journal} {Science}\ }\textbf {\bibinfo {volume} {349}},\ \bibinfo {pages} {711–716} (\bibinfo {year} {2015})}\BibitemShut {NoStop}%
\bibitem [{\citenamefont {Clements}\ \emph {et~al.}(2016)\citenamefont {Clements}, \citenamefont {Humphreys}, \citenamefont {Metcalf}, \citenamefont {Kolthammer},\ and\ \citenamefont {Walsmley}}]{Clements2016}%
  \BibitemOpen
  \bibfield  {author} {\bibinfo {author} {\bibfnamefont {W.~R.}\ \bibnamefont {Clements}}, \bibinfo {author} {\bibfnamefont {P.~C.}\ \bibnamefont {Humphreys}}, \bibinfo {author} {\bibfnamefont {B.~J.}\ \bibnamefont {Metcalf}}, \bibinfo {author} {\bibfnamefont {W.~S.}\ \bibnamefont {Kolthammer}},\ and\ \bibinfo {author} {\bibfnamefont {I.~A.}\ \bibnamefont {Walsmley}},\ }\bibfield  {title} {\bibinfo {title} {Optimal design for universal multiport interferometers},\ }\href {https://doi.org/10.1364/optica.3.001460} {\bibfield  {journal} {\bibinfo  {journal} {Optica}\ }\textbf {\bibinfo {volume} {3}},\ \bibinfo {pages} {1460} (\bibinfo {year} {2016})}\BibitemShut {NoStop}%
\bibitem [{\citenamefont {Reck}\ \emph {et~al.}(1994)\citenamefont {Reck}, \citenamefont {Zeilinger}, \citenamefont {Bernstein},\ and\ \citenamefont {Bertani}}]{Reck1994}%
  \BibitemOpen
  \bibfield  {author} {\bibinfo {author} {\bibfnamefont {M.}~\bibnamefont {Reck}}, \bibinfo {author} {\bibfnamefont {A.}~\bibnamefont {Zeilinger}}, \bibinfo {author} {\bibfnamefont {H.~J.}\ \bibnamefont {Bernstein}},\ and\ \bibinfo {author} {\bibfnamefont {P.}~\bibnamefont {Bertani}},\ }\bibfield  {title} {\bibinfo {title} {Experimental realization of any discrete unitary operator},\ }\href {https://doi.org/10.1103/PhysRevLett.73.58} {\bibfield  {journal} {\bibinfo  {journal} {Phys. Rev. Lett.}\ }\textbf {\bibinfo {volume} {73}},\ \bibinfo {pages} {58} (\bibinfo {year} {1994})}\BibitemShut {NoStop}%
\bibitem [{\citenamefont {Rahman}\ \emph {et~al.}(2023)\citenamefont {Rahman}, \citenamefont {Yang}, \citenamefont {Li}, \citenamefont {Bai},\ and\ \citenamefont {Ozcan}}]{Rahman2023}%
  \BibitemOpen
  \bibfield  {author} {\bibinfo {author} {\bibfnamefont {M.~S.~S.}\ \bibnamefont {Rahman}}, \bibinfo {author} {\bibfnamefont {X.}~\bibnamefont {Yang}}, \bibinfo {author} {\bibfnamefont {J.}~\bibnamefont {Li}}, \bibinfo {author} {\bibfnamefont {B.}~\bibnamefont {Bai}},\ and\ \bibinfo {author} {\bibfnamefont {A.}~\bibnamefont {Ozcan}},\ }\bibfield  {title} {\bibinfo {title} {Universal linear intensity transformations using spatially incoherent diffractive processors},\ }\href {https://doi.org/10.1038/s41377-023-01234-y} {\bibfield  {journal} {\bibinfo  {journal} {Light Sci. Appl.}\ }\textbf {\bibinfo {volume} {12}},\ \bibinfo {pages} {195} (\bibinfo {year} {2023})}\BibitemShut {NoStop}%
\bibitem [{\citenamefont {Piggott}\ \emph {et~al.}(2015)\citenamefont {Piggott}, \citenamefont {Lu}, \citenamefont {Lagoudakis}, \citenamefont {Petykiewicz}, \citenamefont {Babinec},\ and\ \citenamefont {Vučković}}]{Piggott2015}%
  \BibitemOpen
  \bibfield  {author} {\bibinfo {author} {\bibfnamefont {A.~Y.}\ \bibnamefont {Piggott}}, \bibinfo {author} {\bibfnamefont {J.}~\bibnamefont {Lu}}, \bibinfo {author} {\bibfnamefont {K.~G.}\ \bibnamefont {Lagoudakis}}, \bibinfo {author} {\bibfnamefont {J.}~\bibnamefont {Petykiewicz}}, \bibinfo {author} {\bibfnamefont {T.~M.}\ \bibnamefont {Babinec}},\ and\ \bibinfo {author} {\bibfnamefont {J.}~\bibnamefont {Vučković}},\ }\bibfield  {title} {\bibinfo {title} {Inverse design and demonstration of a compact and broadband on-chip wavelength demultiplexer},\ }\href {https://doi.org/10.1038/nphoton.2015.69} {\bibfield  {journal} {\bibinfo  {journal} {Nat. Photonics}\ }\textbf {\bibinfo {volume} {9}},\ \bibinfo {pages} {374–377} (\bibinfo {year} {2015})}\BibitemShut {NoStop}%
\bibitem [{\citenamefont {Hughes}\ \emph {et~al.}(2018)\citenamefont {Hughes}, \citenamefont {Minkov}, \citenamefont {Williamson},\ and\ \citenamefont {Fan}}]{Hughes2018}%
  \BibitemOpen
  \bibfield  {author} {\bibinfo {author} {\bibfnamefont {T.~W.}\ \bibnamefont {Hughes}}, \bibinfo {author} {\bibfnamefont {M.}~\bibnamefont {Minkov}}, \bibinfo {author} {\bibfnamefont {I.~A.~D.}\ \bibnamefont {Williamson}},\ and\ \bibinfo {author} {\bibfnamefont {S.}~\bibnamefont {Fan}},\ }\bibfield  {title} {\bibinfo {title} {Adjoint method and inverse design for nonlinear nanophotonic devices},\ }\href {https://doi.org/10.1021/acsphotonics.8b01522} {\bibfield  {journal} {\bibinfo  {journal} {ACS Photonics}\ }\textbf {\bibinfo {volume} {5}},\ \bibinfo {pages} {4781–4787} (\bibinfo {year} {2018})}\BibitemShut {NoStop}%
\bibitem [{\citenamefont {Khoram}\ \emph {et~al.}(2019)\citenamefont {Khoram}, \citenamefont {Chen}, \citenamefont {Liu}, \citenamefont {Ying}, \citenamefont {Wang}, \citenamefont {Yuan},\ and\ \citenamefont {Yu}}]{Khoram2019}%
  \BibitemOpen
  \bibfield  {author} {\bibinfo {author} {\bibfnamefont {E.}~\bibnamefont {Khoram}}, \bibinfo {author} {\bibfnamefont {A.}~\bibnamefont {Chen}}, \bibinfo {author} {\bibfnamefont {D.}~\bibnamefont {Liu}}, \bibinfo {author} {\bibfnamefont {L.}~\bibnamefont {Ying}}, \bibinfo {author} {\bibfnamefont {Q.}~\bibnamefont {Wang}}, \bibinfo {author} {\bibfnamefont {M.}~\bibnamefont {Yuan}},\ and\ \bibinfo {author} {\bibfnamefont {Z.}~\bibnamefont {Yu}},\ }\bibfield  {title} {\bibinfo {title} {Nanophotonic media for artificial neural inference},\ }\href {https://doi.org/10.1364/prj.7.000823} {\bibfield  {journal} {\bibinfo  {journal} {Photon. Res.}\ }\textbf {\bibinfo {volume} {7}},\ \bibinfo {pages} {823} (\bibinfo {year} {2019})}\BibitemShut {NoStop}%
\bibitem [{\citenamefont {Lin}\ \emph {et~al.}(2019)\citenamefont {Lin}, \citenamefont {Su}, \citenamefont {Wang}, \citenamefont {Chen}, \citenamefont {Chung}, \citenamefont {Chen}, \citenamefont {Kuo}, \citenamefont {Chen}, \citenamefont {Chen}, \citenamefont {Huang}, \citenamefont {Wang}, \citenamefont {Chu}, \citenamefont {Wu}, \citenamefont {Li}, \citenamefont {Wang}, \citenamefont {Zhu},\ and\ \citenamefont {Tsai}}]{Lin2019}%
  \BibitemOpen
  \bibfield  {author} {\bibinfo {author} {\bibfnamefont {R.~J.}\ \bibnamefont {Lin}}, \bibinfo {author} {\bibfnamefont {V.-C.}\ \bibnamefont {Su}}, \bibinfo {author} {\bibfnamefont {S.}~\bibnamefont {Wang}}, \bibinfo {author} {\bibfnamefont {M.~K.}\ \bibnamefont {Chen}}, \bibinfo {author} {\bibfnamefont {T.~L.}\ \bibnamefont {Chung}}, \bibinfo {author} {\bibfnamefont {Y.~H.}\ \bibnamefont {Chen}}, \bibinfo {author} {\bibfnamefont {H.~Y.}\ \bibnamefont {Kuo}}, \bibinfo {author} {\bibfnamefont {J.-W.}\ \bibnamefont {Chen}}, \bibinfo {author} {\bibfnamefont {J.}~\bibnamefont {Chen}}, \bibinfo {author} {\bibfnamefont {Y.-T.}\ \bibnamefont {Huang}}, \bibinfo {author} {\bibfnamefont {J.-H.}\ \bibnamefont {Wang}}, \bibinfo {author} {\bibfnamefont {C.~H.}\ \bibnamefont {Chu}}, \bibinfo {author} {\bibfnamefont {P.~C.}\ \bibnamefont {Wu}}, \bibinfo {author} {\bibfnamefont {T.}~\bibnamefont {Li}}, \bibinfo {author} {\bibfnamefont {Z.}~\bibnamefont {Wang}}, \bibinfo {author} {\bibfnamefont {S.}~\bibnamefont {Zhu}},\ and\
  \bibinfo {author} {\bibfnamefont {D.~P.}\ \bibnamefont {Tsai}},\ }\bibfield  {title} {\bibinfo {title} {Achromatic metalens array for full-colour light-field imaging},\ }\href {https://doi.org/10.1038/s41565-018-0347-0} {\bibfield  {journal} {\bibinfo  {journal} {Nat. Nanotechnol.}\ }\textbf {\bibinfo {volume} {14}},\ \bibinfo {pages} {227–231} (\bibinfo {year} {2019})}\BibitemShut {NoStop}%
\bibitem [{\citenamefont {Jaynes}(1957)}]{Jaynes1957}%
  \BibitemOpen
  \bibfield  {author} {\bibinfo {author} {\bibfnamefont {E.~T.}\ \bibnamefont {Jaynes}},\ }\bibfield  {title} {\bibinfo {title} {Information theory and statistical mechanics},\ }\href {https://doi.org/10.1103/PhysRev.106.620} {\bibfield  {journal} {\bibinfo  {journal} {Phys. Rev.}\ }\textbf {\bibinfo {volume} {106}},\ \bibinfo {pages} {620} (\bibinfo {year} {1957})}\BibitemShut {NoStop}%
\bibitem [{\citenamefont {Han}\ \emph {et~al.}(2015)\citenamefont {Han}, \citenamefont {Pool}, \citenamefont {Tran},\ and\ \citenamefont {Dally}}]{Han2015}%
  \BibitemOpen
  \bibfield  {author} {\bibinfo {author} {\bibfnamefont {S.}~\bibnamefont {Han}}, \bibinfo {author} {\bibfnamefont {J.}~\bibnamefont {Pool}}, \bibinfo {author} {\bibfnamefont {J.}~\bibnamefont {Tran}},\ and\ \bibinfo {author} {\bibfnamefont {W.}~\bibnamefont {Dally}},\ }\bibfield  {title} {\bibinfo {title} {Learning both weights and connections for efficient neural network},\ }in\ \href {https://proceedings.neurips.cc/paper_files/paper/2015/file/ae0eb3eed39d2bcef4622b2499a05fe6-Paper.pdf} {\emph {\bibinfo {booktitle} {Advances in Neural Information Processing Systems}}},\ Vol.~\bibinfo {volume} {28},\ \bibinfo {editor} {edited by\ \bibinfo {editor} {\bibfnamefont {C.}~\bibnamefont {Cortes}}, \bibinfo {editor} {\bibfnamefont {N.}~\bibnamefont {Lawrence}}, \bibinfo {editor} {\bibfnamefont {D.}~\bibnamefont {Lee}}, \bibinfo {editor} {\bibfnamefont {M.}~\bibnamefont {Sugiyama}},\ and\ \bibinfo {editor} {\bibfnamefont {R.}~\bibnamefont {Garnett}}}\ (\bibinfo  {publisher} {Curran Associates, Inc.},\ \bibinfo {year}
  {2015})\BibitemShut {NoStop}%
\bibitem [{\citenamefont {Jia}\ \emph {et~al.}(2021)\citenamefont {Jia}, \citenamefont {Zhu}, \citenamefont {Li}, \citenamefont {Tang},\ and\ \citenamefont {Zhou}}]{Jia2021}%
  \BibitemOpen
  \bibfield  {author} {\bibinfo {author} {\bibfnamefont {X.}~\bibnamefont {Jia}}, \bibinfo {author} {\bibfnamefont {C.}~\bibnamefont {Zhu}}, \bibinfo {author} {\bibfnamefont {M.}~\bibnamefont {Li}}, \bibinfo {author} {\bibfnamefont {W.}~\bibnamefont {Tang}},\ and\ \bibinfo {author} {\bibfnamefont {W.}~\bibnamefont {Zhou}},\ }\bibfield  {title} {\bibinfo {title} {Llvip: A visible-infrared paired dataset for low-light vision},\ }in\ \href@noop {} {\emph {\bibinfo {booktitle} {Proceedings of the IEEE/CVF International Conference on Computer Vision (ICCV) Workshops}}}\ (\bibinfo {year} {2021})\ pp.\ \bibinfo {pages} {3496--3504}\BibitemShut {NoStop}%
\bibitem [{\citenamefont {Redmon}\ and\ \citenamefont {Farhadi}(2018)}]{yolov3}%
  \BibitemOpen
  \bibfield  {author} {\bibinfo {author} {\bibfnamefont {J.}~\bibnamefont {Redmon}}\ and\ \bibinfo {author} {\bibfnamefont {A.}~\bibnamefont {Farhadi}},\ }\bibfield  {title} {\bibinfo {title} {Yolov3: An incremental improvement},\ }\href@noop {} {\bibfield  {journal} {\bibinfo  {journal} {arXiv}\ } (\bibinfo {year} {2018})}\BibitemShut {NoStop}%
\bibitem [{\citenamefont {Mehta}\ and\ \citenamefont {Wolf}(1964)}]{Mehta1964}%
  \BibitemOpen
  \bibfield  {author} {\bibinfo {author} {\bibfnamefont {C.~L.}\ \bibnamefont {Mehta}}\ and\ \bibinfo {author} {\bibfnamefont {E.}~\bibnamefont {Wolf}},\ }\bibfield  {title} {\bibinfo {title} {Coherence properties of blackbody radiation. i. correlation tensors of the classical field},\ }\href {https://doi.org/10.1103/PhysRev.134.A1143} {\bibfield  {journal} {\bibinfo  {journal} {Phys. Rev.}\ }\textbf {\bibinfo {volume} {134}},\ \bibinfo {pages} {A1143} (\bibinfo {year} {1964})}\BibitemShut {NoStop}%
\bibitem [{\citenamefont {Khorasaninejad}\ and\ \citenamefont {Capasso}(2017)}]{Khorasaninejad2017}%
  \BibitemOpen
  \bibfield  {author} {\bibinfo {author} {\bibfnamefont {M.}~\bibnamefont {Khorasaninejad}}\ and\ \bibinfo {author} {\bibfnamefont {F.}~\bibnamefont {Capasso}},\ }\bibfield  {title} {\bibinfo {title} {Metalenses: Versatile multifunctional photonic components},\ }\href {https://doi.org/10.1126/science.aam8100} {\bibfield  {journal} {\bibinfo  {journal} {Science}\ }\textbf {\bibinfo {volume} {358}},\ \bibinfo {pages} {1146} (\bibinfo {year} {2017})}\BibitemShut {NoStop}%
\end{thebibliography}%

\begin{widetext}

\section{Methods}

\subsection{Definition of the block-wise matrices}
In general, a discrete Fourier transform (DFT) tensor in a rectangular domain ($N_x \times N_y$) is defined as
\begin{equation}
    \qty[P_\mathrm{F}(N_x, N_y)]_{k,l,m,n} \equiv \frac{1}{\sqrt{N_x N_y}}\exp\qty[2\pi i\qty(\frac{km}{N_x} + \frac{ln}{N_y}) ],
\end{equation}
where $(k,l)$ and $(m,n)$ are the 2D coordinates satisfying $0\leq k, m < N_x$ and $0\leq l, n < N_y$. Using this definition as a building block, the block-wise DFT tensor can be written as 
    \begin{equation}
    \qty[P_\mathrm{BF}]_{k,l,m,n} \equiv \sum_{X, Y} \qty[P_\mathrm{F}(L_x, L_y)]_{k-x_0,l-y_0,m-x_0,n-y_0} I_X(k)I_Y(l)I_X(m)I_Y(n),
    \end{equation}
where $X = \qty{k\in\mathbb{Z}: x_0 \leq k < x_0 + L_x}$ is iterated over disjoint subsets of integer range $0\leq k < N_x$ slicing the 2D domain into columns, $Y$ defined in the same manner, and $I_X(k)$ is the indicator function that returns 1 if $k\in X$ and otherwise 0. Reshaping the 2D indices into flattened 1D indices, $(k, l) \leftrightarrow \alpha$ and $(m, n) \leftrightarrow \beta$, the unitary matrices $\qty[P_\mathrm{F}]_{\alpha\beta}$ and $\qty[P_\mathrm{BF}]_{\alpha\beta}$ in the main text can be derived using $(k_\alpha, l_\alpha)$ and $(m_\beta, n_\beta)$ which are the quotient-remainder pairs of integers $\alpha$ and $\beta$ with $N_y$, respectively.

\subsection{Diffractive optics}
Based on the Rayleigh-Sommerfeld diffraction integral, the spatial evolution of a scalar electromagnetic wave along $z$-direction can be described as
\begin{equation}
    E(x, y, z_2) = \frac{1}{i\lambda} \int \dd x' \dd y' \frac{e^{ik_0 R}}{R} \frac{z_2-z_1}{R} \qty(1+\frac{i}{k_0 R}) E(x', y', z_1),
\end{equation}
where $\lambda$ and $k_0=2\pi/\lambda$ are a free-space wavelength and the corresponding wave number, respectively, and $R = \qty[(x-x')^2+(y-y')^2 + (z_2-z_1)^2]^{1/2}$ is the distance between the source $(x',y')$ and observation $(x,y)$ points at $z=z_2$ and $z_1$ planes, respectively. Sampling the continuous electric fields with rectangular basis functions as,
\begin{align}
    E(x', y',z_1) &\approx \sum_{k,l} E_{k,l} \mathrm{rect}\qty(\frac{x'-k\Delta x}{\Delta x})\mathrm{rect}\qty(\frac{y'-l\Delta x}{\Delta x}), \\
    E(x, y,z_2) &\approx \sum_{m,n} E_{m,n} \mathrm{rect}\qty(\frac{x-m\Delta x}{\Delta x})\mathrm{rect}\qty(\frac{y-n\Delta x}{\Delta x}), 
\end{align}
where $\mathrm{rect}(a) \equiv 1$ if $|a|<1/2$ and elsewhere 0, a discretized numerical linear relationship can be derived: 
\begin{equation}
    E_{m,n} = G_{m,n}^{k,l}E_{k,l},
\end{equation} 
where
\begin{equation}
    G_{m,n}^{k,l} = \frac{1}{i\lambda}\int_{(k-1/2)\Delta x}^{(k+1/2)\Delta x} \dd x' \int_{(l-1/2)\Delta x}^{(l+1/2)\Delta x} \dd y' \frac{(z_2-z_1) \exp(ik_0 R_{m,n}^{k,l})}{(R_{m,n}^{k,l})^2} \qty(1+\frac{i}{k_0 R_{m,n}^{k,l}}) 
\end{equation}
and $R_{m,n}^{k,l}(z_2-z_1) = [\Delta x^2(k-m)^2 + \Delta x^2(l-n)^2 + (z_2-z_1)^2]^{1/2}$. On top of that, lenses and metalenses in the main text are assumed to be infinitesimally thin and therefore lead to a point-by-point local phase jump, which can be described by $E_{m,n}(z=z_0^+) = \Phi_{m,n} E_{m,n}(z=z_0^-)$, where $z=z_0$ is the location of the lens. By multiplying these transfer relationships alternatively through the lens array, one can obtain the input-output relation of the entire optical system as 
\begin{equation}
    E_{\alpha_L}^\mathrm{(out)} = P_{\alpha_L}^{\alpha_0}E_{\alpha_0}= \qty[G_{\alpha_L}^{\alpha_{L-1}}  \Phi_{\alpha_{L-1}}G_{\alpha_{L-1}}^{\alpha_{L-2}}  \Phi_{\alpha_{L-2}}\cdots G_{\alpha_{1}}^{\alpha_{0}}]E_{\alpha_0}^\mathrm{(in)},
\end{equation}
where $\alpha_l$ for $0\leq l \leq L$ is the flattened 1D index on planes $z=z_l$, including input ($z=z_0$) and output ($z=z_L$) planes.

\end{widetext}

\section{Data availability}
Data that support the plots within this paper and other findings of this study are available from the corresponding author upon reasonable request. Source data are provided with this paper.

\section{Code availability}
All codes are available at GitHub.

\section{Acknowledgements}
The work was supported by the Army Research Office through a Multidisciplinary University Research Initiative program (Grant No. W911NF-22-2-0111).

\section{Author Contributions}
J.K. developed the theory, performed the numerical simulations, and wrote the first manuscript. Z.Y. and N.Y. conceived the idea and supervised the research. All authors discussed the result, edited the manuscript, and approved the content.

\section{Competing Interests}
The authors have no conflicts of interest to declare.

\section{Additional information}
\subsection{Corresponding authors}
Correspondence to Jungmin Kim or Zongfu Yu.

\end{document}



\title{Supplementary Information: Compute-first optical detection for noise-resilient visual perception}

\author{Jungmin Kim}
\email{jmkim93@gmail.com}
\affiliation{Department of Electrical and Computer Engineering, University of Wisconsin-Madison, Madison, WI 53706, USA}

\author{Nanfang Yu}
\affiliation{Department of Applied Physics and Applied Mathematics, Columbia University, New York, NY 10027, USA}

\author{Zongfu Yu}%
\email{zyu54@wisc.edu}
\affiliation{Department of Electrical and Computer Engineering, University of Wisconsin-Madison, Madison, WI 53706, USA}

\date{\today}

\maketitle

\listoffigures
\listoftables
\tableofcontents

\newpage 

\begin{table}[]
    \centering
    \begin{tabular}{ll|rrr}
     \hline  \hline
    Layer & & \multicolumn{3}{c}{Shape} \\
          &  &  Model 1 & Model 2 & Model 3 \\
         \hline \hline
       \textbf{Optical}   & Unitary &  \multicolumn{3}{c}{$(784,784)\in\mathbb{C}$}  \\
     &  Detection & \multicolumn{3}{c}{$(784)\in\mathbb{R}$} \\
          & add G.N., P.N.\\
     \hline
     Pooling &  & \multicolumn{3}{c}{$(784,N_\mathrm{seg}^2)$}  \\
       \hline
       \textbf{Digital}  & BN  \\
       & Linear & $(N_\mathrm{seg}^2,300)$ & $(N_\mathrm{seg}^2,100)$ & $(N_\mathrm{seg}^2,300)$ \\
       & GELU \\
       & Linear & $(300,200)$ &$(100, 40)$ & $(300,250)$ \\
       & GELU \\
       & Linear & $(200,50)$ &$(40, 10)$ & $(250, 200)$ \\
       & GELU \\
       & Linear & $(50,10)$ & NA & $(200,50)$ \\
       & GELU  \\
       & Linear & NA & NA & $(50, 10)$ \\
       & Softmax & \multicolumn{3}{c}{Class probability (10)} \\
       \hline \hline
       
    \end{tabular}
    \caption{Detailed model architecture. G.N., Gaussian noise; P.N., Poisson Noise; BN, Batch normalization; Linear, sequence of dropout with $p=0.2$, fully connected, and BN layers. $N_\mathrm{seg}$ can be 1 to 14 for the networks specified, otherwise 28.}
    \label{tab:network}
\end{table}

\begin{figure}
    \centering
    \includegraphics{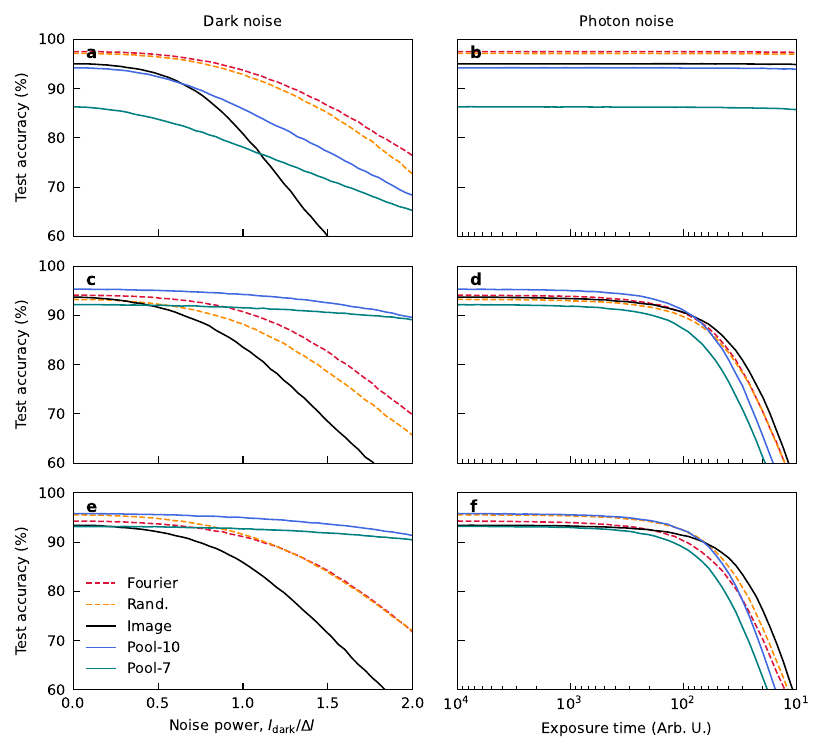}
    \caption{Test accuracy as a function of dark noise power (\textbf{a}, \textbf{c}, \textbf{e}) and shot-noise exposure time (\textbf{b}, \textbf{d}, \textbf{f}), as in Fig. 2. \textbf{a}, \textbf{b}, Sparse input with minimum and maximum value of 0 and 1, respectively, thereby $\Delta I=1$. \textbf{c}, \textbf{d}, Shallower digital network (Model 2 in Table \ref{tab:network}). \textbf{e}, \textbf{f}, Deeper digital network (Model 3 in Table \ref{tab:network}).}
    \label{fig:accuracy}
\end{figure}

\begin{figure}
    \centering
    \includegraphics{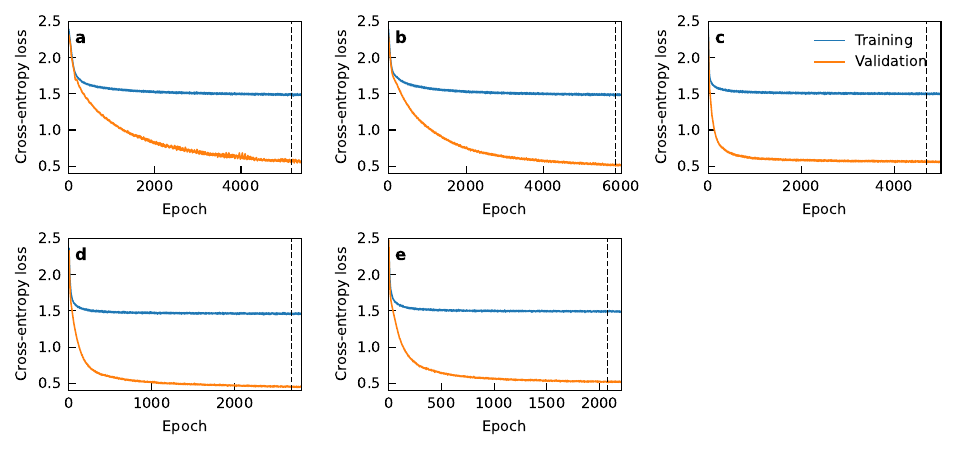}
    \caption{\textbf{Learning curves}. \textbf{a},\textbf{b}, Tandem training of optical and digital networks from the initialization with DFT matrix (\textbf{a}) and a randomly assigned matrix (\textbf{b}). \textbf{c}-\textbf{e}, Digital network-only training with fixed unitary matrices for unprocessed (identity matrix, \textbf{c}) and the block-wise Fourier matrices with 10 (\textbf{d}) and 7 (\textbf{e}) segmentations, as in Fig. 2 in the main text.}
    \label{fig:learningcurve}
\end{figure}

\pagebreak

\section{Incoherent input to the unitary processor}

Assuming an incoherent input $E^\mathrm{(in)}_\alpha(t) = [I_\alpha^\mathrm{(in)}]^{1/2} \exp\qty[i\phi_\alpha(t)]$ with a constant intensity $I_\alpha^\mathrm{(in)}$ and a time-varying phase given by random process $\phi_\alpha(t)$, it is derived that
\begin{align}
    E_\alpha^\mathrm{(out)}(t) &= \sum_\beta P_{\alpha\beta}E_\beta^\mathrm{(in)} = \sum_\beta P_{\alpha\beta} \sqrt{I_\beta^\mathrm{(in)}} \exp[i\phi_\beta(t)] \\
   I_\alpha^\mathrm{(out)}(t)&= \qty|E_\alpha^\mathrm{(out)}(t)|^2  = \sum_{\beta,\beta'} P_{\alpha\beta}^* P_{\alpha\beta'} \sqrt{I_\beta^\mathrm{(in)}I_{\beta'}^\mathrm{(in)}} \exp[i(\phi_{\beta'}-\phi_{\beta})] \\
     \therefore\ev{I_\alpha^\mathrm{(out)}}_t &= \sum_{\beta,\beta'} P_{\alpha\beta}^* P_{\alpha\beta'} \sqrt{I_\beta^\mathrm{(in)}I_{\beta'}^\mathrm{(in)}} \ev{e^{i(\phi_{\beta'}-\phi_{\beta})} }_t \\
    &= \sum_{\beta,\beta'} P_{\alpha\beta}^* P_{\alpha\beta'} \sqrt{I_\beta^\mathrm{(in)}I_{\beta'}^\mathrm{(in)}} \delta_{\beta,\beta'} = \sum_\beta \qty|P_{\alpha\beta}|^2 I_\beta^\mathrm{(in)} = \sum_\beta S_{\alpha\beta}I_\beta^\mathrm{(in)}, \label{eq:intensity}
\end{align}
where $S$ is the transfer matrix for intensity-intensity relation with component $S_{\alpha\beta} = |P_{\alpha\beta}|^2$, and
\begin{equation}
    \ev{f(t)}_t \equiv \lim_{T\rightarrow \infty}\frac{1}{T} \int_0^T \dd t' f(t')
\end{equation}
denotes the average of a time-varying quantity over time, which in this case returns $\delta_{\beta,\beta'}$ due to incoherent random processes $\phi_\beta$ and $\phi_{\beta'}$.

\begin{figure}
    \centering
    \includegraphics{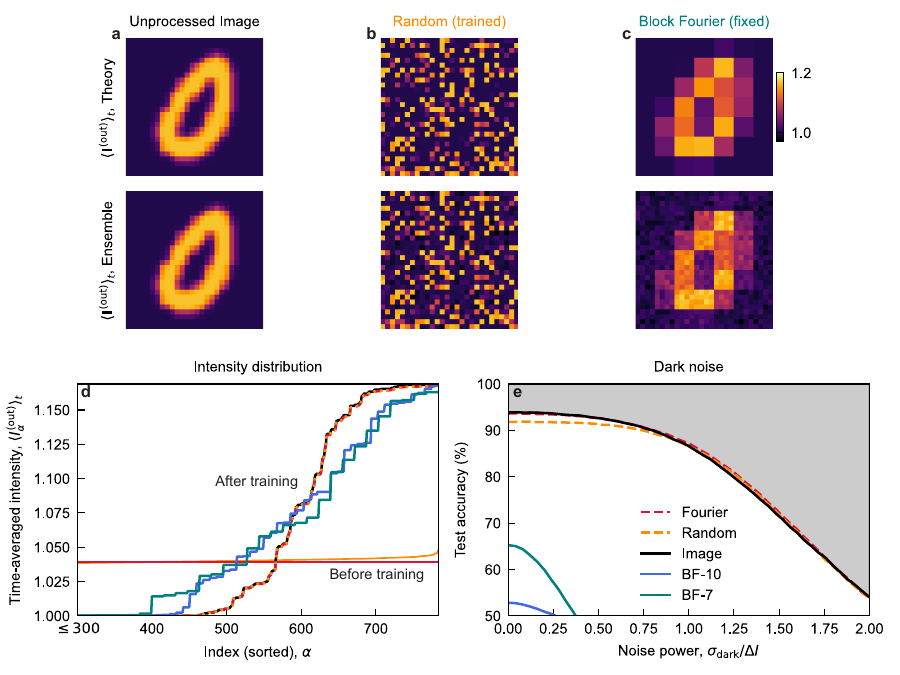}
    \caption{\textbf{Incoherent-input results with unitary matrices}. \textbf{a}-\textbf{c}, Time-averaged output intensity distributions for the unprocessed result (\textbf{a}), the random unitary operation indirectly trained with parameters $P_{\alpha\beta}$ (\textbf{b}), and the block-wise Fourier operation with 7 segments (\textbf{c}), calculated by theoretical average (intensity-intensity relation, top) and ensemble average of random phases (electric field relation, bottom). \textbf{d}, Sorted intensity distributions for various unitary operations, as in Fig. 1 of the main text. \textbf{e}, Test accuracies as a function of dark noise power as in Fig. 1 of the main text.}
    \label{sfig:incoherent}
\end{figure}

Let's suppose that $P$ still remains unitary. Then, by definition $P^\dagger P = I$,
\begin{equation}
    \qty[P^\dagger P]_{\alpha\alpha} = \sum_\beta (P^\dagger)_{\alpha\beta}P_{\beta\alpha} = \sum_\beta (P^*)_{\beta\alpha}P_{\beta\alpha} = \sum_\beta |P_{\beta\alpha}|^2 = \sum_\beta S_{\beta\alpha} = 1
\end{equation}
and similarly $\sum_\beta S_{\alpha\beta}=1$, which means that $S$ is a bistochastic matrix with every column and row summing up to 1. Under this unitary assumption, therefore, the intensity relationship in Eq. \eqref{eq:intensity} means that output time-averaged intensity is the weighted average of input intensity distributions, satisfying $\min(\vb I^\mathrm{(in)}) \leq\ev*{I_\alpha^\mathrm{(out)}}_t \leq \max( \vb I^\mathrm{(in)})$ for any $\alpha$.

From this derivation, it is evident that a unitary constraint does not apply to incoherent wave input, as it fails to concentrate the input intensity. Instead, the intensity diffuses in a manner that reduces the intensity contrast. In Fig. S3, the results for the same setting as in main Fig. 1 are presented, but with an incoherent random phase. As anticipated, Figures S3a-S3d illustrate that the output intensity cannot surpass the maximum input intensity. As a result, the ideal image with the identity matrix emerges as the scenario with the highest intensity contrast. 

Specifically, the training of a unitary matrix leads to the imitation (red and orange dashed lines) of the input intensity distribution (black line) through tailored diffusion, as depicted in Fig. S3d. This is in comparison to the pre-training distributions (red and orange solid lines). Fig. S3e also shows that the ideal imaging is the theoretical upper bound of the noise-robust image recognition, evidenced that trained unitary operations (red and orange dashed lines) converge towards the ideal imaging case. In sharp contrast, the block-wise Fourier operations average out the input intensities at each block as shown in Fig. S3c. Furthermore, the pooling here gives rise to a significant loss of information, which results in extremely low test accuracy, as indicated in Fig. S3e.

\section{Incoherent focusing}

Similar to the block-wise Fourier operation for the unitary and coherent system in the main text, a block-wise intensity-focusing tensor with focusing efficiency $\eta$ for incoherent signals can be defined as
\begin{equation}
\qty[S_\mathrm{focus}(\eta)]_{k,l,m,n} \equiv \sum_{X,Y} \qty{\eta\delta_{k,x_0}\delta_{l,y_0}+ \frac{1-\eta}{L_x L_y-1} \qty[I_X(k)I_Y(l)-\delta_{k,x_0}\delta_{l,y_0}]}I_X(m)I_Y(n),
\end{equation}
which focuses a certain portion ($\eta$) of energy of each input signal into a single output in the corresponding block and uniformly diffuses the remaining ($1-\eta$) part to the other output waveguides. $\qty[S_\mathrm{focus}]_{\alpha\beta}$ can be also obtained by 1D reshaping of indices.

By optimizing the digital network $\vb y = g(\vb x)$ with the incoherently detected input $x_\alpha =  \langle I_\alpha^\mathrm{(out)}\rangle_t + \Delta I_\mathrm{dark} +\Delta I_\mathrm{photon}$ with the optical focusing unit $\langle\vb I^\mathrm{(out)}\rangle_t = S_\mathrm{focus}\vb I^\mathrm{(in)}$, we observe that the compression of optical signals still plays a crucial role in immunity against detection noise as depicted in Fig. S4. For segmentation numbers $N_\mathrm{seg}= 7$ and 10, with two different focusing efficiencies $\eta = 0.5$ and 0.99, every combination of the optical compressive networks exhibits superior noise robustness compared to the reference model without processing. In particular, the focusing efficiency $\eta$ becomes a key parameter for achieving almost zero accuracy drop with an increase in dark noise, similar to the block-wise Fourier operation for coherent signals.

\begin{figure}
    \centering
    \includegraphics{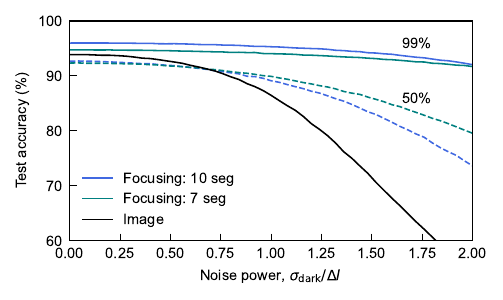}
    \caption{\textbf{Incoherent focusing}. Using block-wise intensity-focusing OPU with the linear intensity relationship, similar noise-robust results can be obtained depending on focusing efficiencies; 99\% (solid coloured lines) or 50\% (dashed lines) of the energy of each waveguide is focused into a single pixel per block. blue, 10 segments; green, 7 segments. }
    \label{fig5}
\end{figure}

\begin{figure}
    \centering
    \includegraphics{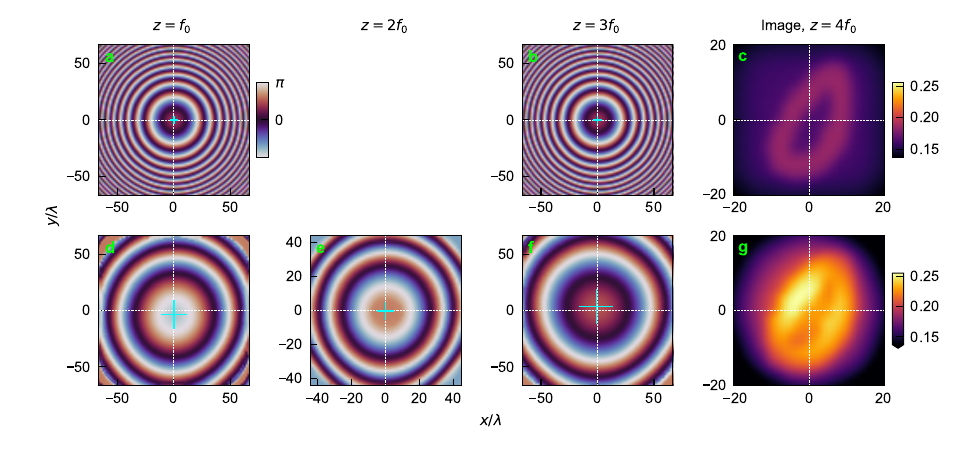}
    \caption{\textbf{Structures of incoherent imaging devices}. \textbf{a}--\textbf{c}, Phase profiles of lenses (\textbf{a}, L1; \textbf{b}, L3) at $z=f_0$ and $3f_0$, and an example image (\textbf{c}) at $z=4f_0$ of the conventional $4f$ system.  \textbf{d}--\textbf{g}, Phase profiles of metalenses (\textbf{d}, ML1; \textbf{e}, ML2; \textbf{f}, ML3) at $z=f_0, 2f_0$ and $3f_0$, and the corresponding image (\textbf{g}) at $z=4f_0$ of the trained meta-imaging system.}
    \label{fig:enter-label}
\end{figure}



\nocite{*}
